\documentclass[preprint]{aastex}

\newcommand{\OI}{[\ion{O}{1}]}
\newcommand{\OIII}{[\ion{O}{3}]}
\newcommand{\SII}{[\ion{S}{2}]}
\newcommand{\NII}{[\ion{N}{2}]}

\newcommand{\OIIIHb}{$\log([$\ion{O}{3}$]/{\rm H}\beta)$}
\newcommand{\Ha}{H$\alpha$}
\newcommand{\Hb}{H$\beta$}
\newcommand{\HeII}{\ion{He}{2}}

\newcommand{\CIII}{\ion{C}{3}}

\newcommand{\SiIII}{\ion{Si}{3}}

\newcommand{\NIII}{\ion{N}{3}}
\newcommand{\FeIII}{[\ion{Fe}{3}]}
\newcommand{\HeIIHb}{\ion{He}{2}$\,\lambda 4686/{\rm H}\beta\,$}
\newcommand{\logHeIIHb}{$\log$(\ion{He}{2}$\,\lambda 4686/{\rm H}\beta\,$)}
\newcommand{\LIR}{${\rm L_{IR}}\,$}
\newcommand{\LFIR}{${\rm L_{FIR}}\,$}

\usepackage{afterpage}

\begin{document}

\title{Theoretical Modeling of Starburst Galaxies}
\author{L.J. Kewley, M.A. Dopita, R. S., Sutherland, C.A. 
Heisler\altaffilmark{1}, J. Trevena}

\begin{abstract}
We have modeled a large sample of infrared starburst galaxies using
both the PEGASE v2.0 and STARBURST99 codes to generate the spectral energy
distribution (SED) of the young star clusters. PEGASE utilizes the Padova
group tracks while STARBURST99 uses the Geneva group tracks, allowing
comparison between the two. We used our MAPPINGS~III code to compute
photoionization models which include a self-consistent treatment of dust
physics and chemical depletion. We use the standard optical diagnostic
diagrams as indicators of the hardness of the EUV radiation field in these
galaxies. These diagnostic diagrams are most sensitive to the spectral index
of the ionizing radiation field in the 1-4 Rydberg region. We find that warm
infrared starburst galaxies contain a relatively hard EUV field in this
region. The PEGASE ionizing stellar continuum is harder in the 1-4 Rydberg 
range than that of STARBURST99.   As the spectrum in this regime is
dominated by emission from Wolf-Rayet (W-R) stars, this difference is
most likely due to the differences in stellar atmosphere models used for
the W-R stars.  The PEGASE models use the \citet{Clegg87}
planetary nebula nuclei (PNN) atmosphere models for the Wolf-Rayet stars 
whereas the STARBURST99 models use the \citet{Schmutz92} Wolf-rayet atmosphere
models.  We believe that the \citet{Schmutz92} atmospheres are more
applicable to the starburst galaxies in our sample, however they do not produce the hard EUV field in the 1-4 Rydberg region required by our observations.
The inclusion of continuum metal blanketing in the models may be one 
solution.  
Supernova remnant (SNR) shock modeling shows that the contribution by
mechanical energy from SNRs to the photoionization models is $\ll 20$\%.
The models presented here are used to derive a new theoretical 
classification scheme for starbursts and AGN galaxies based on the 
optical diagnostic diagrams.

\end{abstract}

\affil{Research School of Astronomy and Astrophysics, Australian National
University} \authoraddr{Private Bag, Weston Creek PO, ACT 2611, Australia}

\altaffiltext{1}{deceased, 28 October 1999}
\maketitle

\section{Introduction}

Observations of starburst galaxies can provide vital insights into the
processes and spectral characteristics of massive star formation regions. In
such regions the physical conditions are similar to those that existed at
the time of collapse and formation of galaxies in the early universe, and
they can also provide an understanding of early galaxy evolution. The
{\em Infrared Astronomical Satellite} (IRAS) made the key discovery of large
numbers of infrared luminous galaxies, similar to those found by 
\citet{Rieke72}. Many of these are dominated by intense star formation %
\citep{Lutz96,Lutz98,Genzel98,Veilleux95,Veilleux99} in which the luminosity of
the young hot stars heats the surrounding dust, producing large amounts of
infrared radiation.

The theoretical tools required to interpret the spectra of such galaxies are
now available. For example, detailed stellar population synthesis models
have been developed for both instantaneous and continuous starbursts and
using these models, one is able to derive parameters such as the starburst
age and metallicity from the continuous spectrum. In such models, the
stellar initial mass function (IMF), star formation rate (SFR) and stellar
atmosphere formulations are all adjustable initial parameters.

The emission line spectrum of the starburst provides constraints on the
physical parameters for the ionized gas and the interstellar medium in
general. In particular, the gas density, temperature and pressure can be 
derived directly from such observations, and the total rates of star
formation can be estimated from the luminosity in the Balmer lines of
hydrogen for the objects without large quantities of dust at least 
({\em eg} \citet{Kennicutt98}).
Using the ionizing UV radiation fields produced by stellar population 
synthesis models in conjunction with detailed self-consistent 
photoionization models such as MAPPINGS~III \citep{Sutherland93} or 
CLOUDY \citep{Ferland98} we can now generate models for any \ion{H}{2} 
region or starburst. In such models it is vital to include a 
self-consistent treatment 
of dust physics and the depletion of various elements out of the gas phase.

Since the nebular emission line spectrum is very sensitive to the
hardness of the ionizing EUV radiation, optical line ratio diagnostic
diagrams provide an important constraint on the shape of the EUV spectrum
and these may also be used to estimate the mean ionization parameter and
metallicity of the galaxies. Such optical diagnostic diagrams were first
proposed by \citet{Baldwin81} to classify galaxies into starburst or AGN
type, since AGN have a much harder ionizing spectrum than hot stars. The
classification scheme was revised by \citet{Osterbrock85} and %
\citet{Veilleux87}, hereafter VO87. These revised diagnostics are used here.
For both schemes, the line diagnostic tools are based on emission-line
intensity ratios which turn out to be particularly sensitive to the 
hardness of the EUV radiation field.

In an earlier paper \citep{Dopita00a}, we theoretically recalibrated the
extragalactic \ion{H}{2} region sequence using these line diagnostic
diagrams and others, in order to separate and quantify the effects of
abundance, ionization parameter and continuous vs. instantanteous burst
models. The theoretical \ion{H}{2}
region models were generated by the MAPPINGS~III  code
which uses as input the EUV fields predicted by the stellar population
synthesis models PEGASE v2.0 \citep{Fioc97} and STARBURST99 \citep{Leitherer99}.
Dust photoelectric heating and the gas-phase depletion of the heavy
elements were treated in a self-consistent manner. This work found that the
high surface brightness isolated extragalactic \ion{H}{2} regions are in
general excited by young clusters of OB stars, and that, in this case, the
ionizing EUV spectra and \ion{H}{2} region emission line spectra predicted 
by the PEGASE and STARBURST99 codes are essentially identical.

For starburst galaxies, in which the starburst has a luminosity comparable
to the luminosity of the host galaxy, the situation is rather different. In 
these objects, intense
star formation is likely to continue over at least a galactic dynamical
timescale, and therefore the assumption of a continuous rather than an
instantaneous burst of star formation would be more accurate. As a 
consequence,
the assumptions which go into the theoretical stellar mass loss formulations
and evolutionary tracks are likely to play a much more important role in the
modeling. Furthermore, for starbursts continued for more than a few Myr,
the Wolf-Rayet (W-R) stars can play an important part in determining both
the intensity and shape of the EUV spectrum.  For the W-R stars, the 
uncertain assumptions made about the stellar lifetimes, wind mass-loss 
rates, the velocity law in the stellar wind, and the atmospheric opacities 
play a critical role in determining the spectral shape and intensity of the 
emergent EUV flux predicted by theory.

In this paper, we present new grids of theoretical models (based on the
assumption of continuous star formation) which again combine \ion{H}{2}
region models generated by the MAPPINGS~III code with
input EUV fields given by the stellar population spectral synthesis models
PEGASE 2 and STARBURST99. These models are
used in conjuction with our large observational data set described in 
\citet{Kewley00a} and \citet{Kewley00b} to place new {\em observational}
constraints on the shape of the EUV ionizing radiation field. Since the two
stellar population spectral synthesis codes provide a wide choice of 
stellar mass loss formulations, evolutionary tracks and stellar atmospheric 
transfer models, they provide strikingly different predictions about the 
shape and intensity of the EUV field as a function of stellar age. In this 
paper, we use these to separate and quantify the effects of the stellar 
atmospheric models and the evolutionary tracks used on the optical 
diagnostic diagrams. In particular, we will show that the models which 
give the hardest EUV spectrum below the \HeII~ionization limit, but 
which have relatively few photons above this limit, provide the best 
empirical fit to the distribution of starburst galaxies on the optical
line ratio diagnostic diagrams. These new observational constraints should 
prove very helpful to theoreticians modeling the late stages of
stellar evolution in massive stars.

This paper is structured as follows.  Our observational comparison sample
is described in Section 2.  The stellar population synthesis models used
to calculate the EUV ionizing radiation field are presented in Section 3.
Our theoretical starburst models were produced using the photoionization
and shock code MAPPINGS~III and are described in Section 4.  Wolf-Rayet 
emission in our sample of starburst galaxies is discussed in Section 5
and the effect of continuum metal opacities on the EUV ionizing continuum
is discussed in Section 6.  A new theoretical classification
scheme for starbursts and AGN is presented in Section 7 and our main
conclusions are summarised in Section 8.

\section{The Observational Comparison Sample}

We have selected a large sample of 285 warm IRAS galaxies covering a wide
range of infrared luminosities. The warm selection criterion ensures that
our sample contains galaxies with either concentrated star formation 
\citep{Armus89,Armus90}, AGN \citep{Miley85}, or in many cases, a
combination of the two.

Our sample has been selected from the catalogue by \citet{Strauss92} and
consists of all objects south of declination $\delta = +10^{\circ}$ with the additional
following criteria;

1. Flux at 60 ${\rm\mu m} > 2.5 \,{\rm Jy}$ with moderate or high quality
detections at 25, 60 and 100 ${\rm \mu m}$

2. Redshift $<$ 8000 km \,s$^{-1}$ for $\log ({\rm L_{FIR}}) < 11$ and $< $ 30000
km \,s$^{-1}$ for $\log ({\rm L_{FIR}}) > 11$

3. Galactic latitude $|b|>$ 15$^{0}$, and declination $\delta <$ +10$^{\circ}$

4. Warm FIR colours; $8>{\rm F}_{60}/{\rm F}_{25}>0.5$ and $2>{\rm F}_{60}/%
{\rm F}_{100}>0.5$

\noindent where ${\rm F_{25}}$,${\rm F_{60}}$ and ${\rm F_{100}}$ are the
IRAS fluxes at 25, 60 and 100 ${\rm \mu m}$ respectively. These selection
criteria ensure that that the galaxies in the sample are well-resolved,
and that the sample has a
large dynamical range in luminosity, so that luminosity dependent effects
can, in principle, be investigated.

High resolution (50 km s$^{-1}$ at ${\rm H}\beta $) optical spectra with
useable S/N ($>3\sigma$ )ratios were obtained for 225 of the galaxies in our sample.
Spectra were obtained in the red and blue wavelength ranges using the Double
Beam Spectrograph on the Mount Stromlo and Siding Springs 2.3m telescope.
These selection criteria and the full details of our observations are
discussed in detail in \citet{Kewley00a} and \citet{Kewley00b}. All 225
objects were classified into AGN, starburst and LINER types using new
theoretical classification lines on the \citet{Veilleux87} (hereafter
VO87) diagnostic diagrams. We found 157 starburst galaxies using the
VO87 diagrams and semi-empirical classification scheme which provide the 
primary observational comparison for the starburst spectral modeling 
presented here.

We believe there is little contamination by obscured AGN in this starburst 
sample for two reasons.  Firstly, in \citet{Kewley00a}, we have shown
that the optical diagnostic diagrams are extremely sensitive to the
presence of an AGN.  An AGN which contributes only 20\% to the optical
emission increases the line ratios sufficiently that it would be 
classified as an AGN.  Secondly, we found very few starburst galaxies with
warm colors ($\sim 4<{\rm F}_{60}/{\rm F}_{25}<8$).  Since warm colors 
usually mean that the galaxy is energetically dominated by an AGN we can 
conclude that the fraction of obscured AGN in our starburst galaxies
must be low.   These results are consistent with the infrared studies of
\citet{Veilleux99} and \citet{Genzel98}, who showed that ultraluminous 
infrared galaxies classified as starbursts also show a lack of an 
energetically important AGN.

\section{Stellar Population Synthesis Models}

The models we have utilized are described in detail in \citet{Dopita00a}, and
we describe them again briefly here. We have used both the PEGASE 2
and STARBURST99 codes to model the starbursts in our sample. The PEGASE 2 
code uses the \citet{Lejeune97} grid of atmospheres covering the entire 
Hertzsprung-Russell diagram (HRD) plus
\citet{Clegg87} planetary nebula nuclei (PNN) atmospheres for stars with 
high effective temperatives
(T $>$ 50000 K) (hereafter known as the Clegg \& Middlemass atmospheres).  
The \citet{Lejeune97} grid (hereafter
called the Lejeune grid) is derived from three sets of
atmosphere calculations, the bulk being the \citet{Kurucz92} models with 
smaller specialized cool star models by \citet{Fluks94} and 
\citet{Bessell89,Bessell91}.  Lejeune, Cuisinier \& Buser incorporated
observational flux corrections into these models for a range of
stellar temperatures, but does not 
alter the $>30$K models of Kurucz.  The STARBURST 99 code also uses the 
plane-parallel atmospheric Lejeune grid.  For
stars with strong winds it offers the choice of the Lejeune
 grid or the \citet{Schmutz92} extended model
atmospheres (hereafter known as the Schmutz atmospheres). The prescription for 
the switch between extended and plane-parallel atmospheres is the same as 
in \citet{Leitherer95}.

We ran models for both the ``standard'' mass loss rates and the 
``enhanced'' mass loss rates described in \citet{Leitherer99}.  We found
that for the line diagnostic ratios used here, the mass loss prescriptions 
agree to within 0.03 dex.  

Both codes follow the theoretical stellar tracks from the Zero Age Main
Sequence (ZAMS) to their final stages. These stages include the
 asymptotic and post-asymptotic giant branch phases for intermediate-mass 
stars. The Padova tracks 
\citep{Bressan93} are used
in PEGASE 2 and the Geneva tracks (\citet{Schaller92}) are used in STARBURST99. 
The Padova tracks
overshoot for masses $m\ge 1\,{\rm M}_{\odot }$ and use a higher ratio of the
overshooting distance to the pressure scale height and down to lower masses
than the Geneva tracks, which include overshooting only above 1.5 ${\rm M}%
_{\odot }$. Both tracks use the OPAL opacities; \citet{Iglesias92} for the
Padova Tracks and \citet{Rogers92} for the Geneva tracks. Both sets of
tracks assume similar mixing lengths. Helium contents of 0.28 and 0.30 are
used for the Padova and Geneva tracks respectively. The Padova tracks have a
higher resolution in mass and time. Clear differences between these two sets
of tracks (at solar metallicity) can be seen by comparing Figure (5) in 
\citet{Schaller92} and Figure (7) in \citet{Bressan93}.

Assuming a standard initial mass function, the choices offered by the two
spectral synthesis modeling codes allows sufficient flexibility to
separately investigate, and to quantify, the effect that either the stellar
atmospheres or the stellar evolutionary tracks have upon the theoretical
starburst model line intensity ratios as a function of age. For
example we can run the STARBURST99 code with either the Lejeune
atmospheres or the Lejeune plus Schmutz atmospheres
to investigate the effect of extended atmospheres, or we could 
compare the PEGASE 2  and STARBURST99  
codes to quantify the effect of the differences
between the Padova and the Geneva tracks and the effect of the
different stellar atmosphere models incorporated into these codes. We
are confident that this
comparison is valid, since in our earlier paper \citep{Dopita00a}, we
compared the ionizing EUV spectra and \ion{H}{2} region emission spectra
predicted by the PEGASE and STARBURST99 codes for zero age clusters, and
found these to be essentially identical.

In the modeling of the starburst emission spectra, we distinguish
between a zero-age {\em instantaneous} star formation case, and {\em 
continuous} starburst models in which a balance between star birth and star
death is set up for all stellar masses which contribute significantly to the
EUV spectrum.  The hydrogen-burning lifetime of massive stars is
approximately $\tau =4.5({\rm M}/40{\rm M}_{\odot })^{-0.43}$ Myr, so in practice this
condition is satisfied for any starburst which lasts longer than about 6
Myr. This gives a dynamical balance between star births and star deaths for
all masses greater than about $20\,{\rm M}_{\odot }$, and is also time enough for
the Wolf Rayet stars to produce their full contribution to the EUV spectrum.

Figure (1) compares predicted solar metallicity EUV spectra produced by
the PEGASE code
which uses the Lejeune stellar atmospheres plus
Clegg \& Middlemass atmospheres for stars with high effective temperatives
(T $>$ 50000K), and the 
STARBURST99 codes which use either the Lejeune stellar
atmospheres grid or the Lejeune atmospheres plus 
Schmutz extended model atmospheres. The PEGASE models used
range from ages of $0-6$ Myr,  and the STARBURST99 models cover ages of $0-8$
Myr. No further evolution in the shape of the EUV spectrum is seen after 6
and 8 Myr for the PEGASE and STARBURST99 models respectively.

After a few Myr of evolution, quite marked differences in the EUV spectrum
develop. The comparison of Figure (1b) and Figure (1c) shows that the 
Schmutz extended atmospheres produce far more ionizing radiation
at frequencies above the \HeII~ionization limit than do the 
Lejeune atmospheres, but the diffences are much less marked
at lower energies. The differences between Figures (1a-c) are
due to a combination of the different evolutionary tracks, and the 
stellar atmosphere models used.  Figures (2) and (3) show the
EUV spectra from PEGASE 2 and STARBURST99 for metallicities 0.2
and 2 $\times$ solar.  We can observe clear differences in the EUV
spectra with increasing metallicity.  In particular, the EUV spectrum
becomes harder for lower age models.  This is expected if high mass
stars are responsible for the EUV field in this regime.  At higher 
metallicities, the high mass stars make a larger contribution to the EUV
radiation field at younger burst ages.  The most likely cause of the
difference between the EUV fields in Figures (1a-c) is to be found in the 
different stellar atmospheric models used for the high mass stars, 
especially for the Wolf-Rayet stars. 

Since it was specifically constructed to model starbursts, the STARBURST99
code uses a more theoretically sophisticated approach to modeling the EUV
spectrum. In the PEGASE 2 models, stars with effective temperatures 
greater than 50000 K (which includes Wolf-Rayet stars) are modeled by the
Clegg \& Middlemass planetary nebulae nuclei (PNN) atmospheres. These stars have
much higher surface gravity than Wolf-Rayet stars, and so we would expect
the atmospheric blanketing to be quite different. In STARBURST99 code, 
stars with strong stellar winds (which includes Wolf-Rayet stars)
are modeled by the Schmutz Wolf-Rayet atmospheres.  These 
include H and He opacities, but do not include heavy element opacities.

The EUV spectrum emergent from a W-R model atmosphere depends critically on
the fraction of ionizing photons which have been used up to maintain the
ionization of the W-R wind region. This is determined by the size of the
emission measure of the atmosphere; $\int n_{{\rm e}}^{2}dr.$ This parameter
is proportional to the the product $\left( \stackrel{.}{M}/v_{\infty
}\right) ^{2}R_{*}^{-3}$, where  $\stackrel{.}{M}$ is the mass-loss rate, $
v_{\infty }$ the terminal velocity of the wind, and $R_{*}$ is the
photospheric radius of the star. This product is the \citet{Schmutz89}
density parameter. Models with the same density parameter display very
similar emission line equivalent widths, but the total scaling in luminosity
depends on $R_{*}^{2}.$ The density parameter can also be expressed in terms
of a ``transformed radius'', $R_{{\rm t}}$:

\[
R_{{\rm t}}=R_{*}\left[ \frac{v_{\infty }\stackrel{.}{M}_{ref}}{v_{ref}%
\stackrel{.}{M}}\right] ^{2/3}
\]
where $v_{ref}$ is a (normalizing) reference velocity, and $\stackrel{.}{M}%
_{ref}$ is a (normalizing) reference mass-loss rate. Again, models with
similar values of the transformed radius give similar spectra. Stars
which use a greater fraction of their EUV photons in maintaining the 
photoionization of their extended atmospheres would show a lower-intensity,
harder EUV spectrum below the \ion{He}{2} ionization edge, and would be 
expected to exhibit more atmospheric blanketing by heavy elements.

\section{Starburst Modeling}

To model the starburst spectrum as a function of age of the exciting stars,
metallicity and ionization parameter, we input the ionizing spectrum from
the both continuous and instantaneous PEGASE and STARBURST99 models into the
MAPPINGS~III code. The photoionization modeling carried out with MAPPINGS~III
for this analysis is described in \citet{Dopita00a} and is described briefly
here. We computed plane parallel, isobaric models with electron density
of 350~${\rm cm^{-3}}$, which is the average electron density of the individual (frequently unresolved) \ion{H}{2} regions within the 1~kpc slit aperture
extracted for the starbursts in our sample.  The electron density
was found using the flux in the \SII~$\lambda 6716$ and \SII~$\lambda 6731$
forbidden lines from our spectra in conjunction
with a 5 level model atom using MAPPINGS~III.  Electron densities for
the galaxies in our sample can be found in \citet{Kewley00a}. 

The ionization parameter $q$ (cm~s$^{-1}$) is defined on the inner
boundary of the nebula, that is, the boundary nearest the exciting star.
This dimensional ionization parameter can be readily transformed to the more
commonly used dimensionless ionization parameter through the identity ${\cal 
U}\equiv q/c.$

Dust physics is treated explicitly through the absorption of the radiation
field on grains, grain charging and photoelectric heating by the grains. We
do not yet calculate the re-emission spectrum of dust in the IR but this is
currently being implemented.  
The dust model consists of silicate grains (100\AA\ $<a<$ 1000\AA ) and small amorphous organic grains (10\AA\ $<a<$
100\AA ), with a size distribution following a \cite{Mathis77} power-law
and spherical geometry.  The range of sizes is chosen so as to give depletion factors relative to
solar similar to those observed by UV interstellar absorption measurements
of stars seen through warm diffuse clouds in the local interstellar medium
{\em eg} \cite{Jenkins87}. The MAPPINGS~III dust model also provides the 
observed absorption per hydrogen atom for solar metallicity \citep{Bohlin78}:
\[
\frac{N(H)}{E_{B-V}}\sim 5.9\mbox{\rm   x}10^{21}
\mbox{\rm   cm$^{-2}$
mag$^{-1}$ }
\].

Photoelectric yields are found using a more conservative yield curve of the 
same form as \citet{Draine87}.  Photoelectric grain currents are found using
\citet{Draine78} using the dust absorption data from \citet{Laor93}.
Collisions with electrons and protons are considered assuming the standard
``sticking'' coefficients, following \citet{Draine78} and \citet{Draine87}.
More details of the dust physics in MAPPINGS~III can be found in \citet{Dopita00b}.

The undepleted solar abundances are assumed to be those of 
\citet{Anders89}; these abundances and the depletion factors adopted for
each element in the starburst modeling are shown in Table 1. For
non-solar metallicities we assume that both the dust model and the depletion
factors are unchanged, since we have no way of estimating what they may be
otherwise.

All elements except nitrogen and helium are taken to be primary
nucleosynthesis elements. It is known that this assumption may be incorrect
in systems where the time history of star formation in the galaxy is
different, or where galactic winds are important. For example, the O/Fe
ratio is different from its solar value in both the LMC and the SMC 
\citep{Russell92}. Again, we are forced use the simplest assumptions
possible in the absence of a more detailed understanding of the chemical
evolution of starburst galaxies.

For helium, we assume a primary nucleosynthesis component in addition to the
primordial value derived from \cite{Russell92}. This primary component is
matched empirically to provide the observed abundances at SMC, LMC and solar
abundances; \cite{Anders89}, \cite{Russell92}. By number, the He/H ratio is:
\[
\frac{{\rm He}}{{\rm H}}=0.081+0.026(Z/Z_{\odot })
\]
Nitrogen is assumed to be a secondary nucleosynthesis element above
metallicities of $0.23$ solar, but as a primary nucleosynthesis element at
lower metallicities. This is an empirical fit to the observed behaviour of
the N/O ratio in \ion{H}{2} regions (van Zee, Haynes \& Salzer, 1997). By
number:

\begin{eqnarray*}
\log ({\rm N}/{\rm H}) &=&-4.57+\log (Z/Z_{\odot })
\mbox{\rm{  ;  }$\log
(Z/Z_{\odot })\leq -0.63$} \\
\log ({\rm N}/{\rm H}) &=&-3.94+2\log (Z/Z_{\odot })
\mbox{\rm  { ;
otherwise}}
\end{eqnarray*}

The abundances and depletion factors for each element used can be found in
Table \ref{table1}.

\placetable{table1}

The ionization parameter was varied from $q=5\times 10^{6}\,{\rm cm\,s}^{-1}$
to $q=3\times 10^{8}\,{\rm cm\,s}^{-1}$, and the metallicities varied from
0.01 to 3 solar for PEGASE and 0.05 to 2 solar for STARBURST99. The
metallicity values used are restricted by the stellar tracks used by the
population synthesis models.

\subsection{Instantaneous Models}

Figure (4) shows that, due to the absence of Wolf-Rayet stars, the
use of similar zero-age Main Sequences, and the use of identical model
atmospheres for massive stars, the shape of the EUV spectrum for both 
the PEGASE 2 and STARBURST99 models is almost identical for 
instantaneous burst models. As a consequence, they produce almost 
identical optical line ratios in the photoionization models. The results 
from these models are compared 
with the observational data set on the VO87 line ratio diagnostic diagrams 
in Figures (4), (5) and (6). Note that the major problem seen with these 
models is that many starburst galaxies are found in a region lying 
above and to the right of the ``fold'' in the ionization parameter : 
metallicity surface. This presents a problem, since these points lie in 
a ``forbidden zone'' of line ratio space which cannot be reached by 
any combination of metallicity or ionization parameter.
The only way to make models which fall into this region of the the
diagnostic diagrams is either to mix in another type of excitation {\em i.e.}
shocks or a power-law ionizing radiation field or simply to use a harder
EUV ionizing spectrum, particularly in the 1-4 Rydberg region.

In any event, we should not be too surprised that the instantaneous models
do not provide a very good fit to the observed starbursts, since many of
these objects are seen in merging pairs of galaxies, and in this case we
would theoretically expect star formation to continue over a galactic
dynamical timescale. Thus massive clusters associated with individual 
(usually unresolved) \ion{H}{2} regions should have a wide variety of ages.
Direct evidence for continuous star formation, or older \ion{H}{2} regions,
is seen in many of our starburst spectra.   These
show either a low equivalent width in H$\beta$ (which often indicates
dilution by an older stellar population) or clear evidence of 
H$\beta $ absorption in the stellar continuum.  Both of these indicate that 
star formation has continued over at least several Myr \citep{Gonzalez99}.

In \citet{Dopita00a}, we showed that the
line ratio usually used for measuring the ionization parameter; [\ion{O}{3}
]~$\lambda$5007 / [\ion{O}{2}]~$\lambda \lambda$3726,9, is indeed a good
diagnostic and that
the [\ion{N}{2}]~$\lambda$ 6584 / [\ion{O}{2}]~$\lambda \lambda$3726,9
ratio gives the best 
diagnostic of abundance, as it is monotonic between 0.1 and over 3.0
times solar metallicity.  The wavelength range of our spectra was not 
sufficient for us to observe [\ion{O}{2}], however we present the
grids of the instantaneous models for the 
[\ion{N}{2}]~$\lambda \lambda$6548,84 / [\ion{O}{2}]~$\lambda 
\lambda$3726,9~{\em vs. }[\ion{O}{3}]~$\lambda$5007 /H$\beta$ diagram and
the [\ion{N}{2}]~$\lambda \lambda$6548,84/[\ion{O}{2}]~$\lambda
 \lambda$3726,9~{\em vs. } [\ion{O}{3}]~$\lambda$5007/[\ion{O}{2}]~
$\lambda \lambda$3726,9 in Figures (7) and (8) for the use of the
astronomical community.

\subsection{Continuous Models}

As we have seen, when star formation continues over several Myr, the
ionizing spectrum evolves until a dynamic balance between stellar births and
stellar deaths has been set up for all intial stellar masses which are
important in producing EUV photons. As Figure (1) shows, this occurs after 6
and 8 Myr for the PEGASE and STARBURST99 models respectively, and so these
cluster ages were assumed for the continuous star formation models presented
in this section.

The results for the continuous models shown on the VO87 line ratio 
diagnostic diagrams are presented in Figures (9-11) for the 
PEGASE 2 models with (a) the Lejeune plus Clegg \& Middlemass PNN model 
atmospheres, (b) the STARBURST99 models with the Lejeune plus Schmutz model 
atmospheres and (c) the STARBURST99 models with the Lejeune atmospheres 
respectively.

Let us first consider the two sets of STARBURST99 models. The differences
seen here simply reflect the differences in the stellar atmospheric models.
The Lejeune plus Schmutz model atmospheres show little change in spectral
slope between 1 and 3 Ryd as a function of cluster age. It is therefore not
surprising that these continuous star formation models give very similar
results to the zero age instantaneous models of Figures (4 - 8). In these
continuous star formation models the W-R stars provide a radiation
field of appreciable strength above the \HeII~ionization limit (eg. 4-8 Ryd
in Figures 1-3). In this
case, we might expect to detect \HeII~lines in the optical spectrum.
The detection of a nebular \HeII $\,\lambda 4686$ line would provide an
important observational diagnostic in support of the Schmutz extended
atmospheric modeling and is discussed in the following section.

These models with Schmutz extended atmospheres have exactly the same 
difficulty in explaining the position of
the observed points as do the instantaneous models, in that the model grid
fails to overlap about half of the observed points, indicating the need for
a harder ionizing spectrum than these tracks and atmospheres provide.

The second of the STARBURST99 grids, for continuous starbursts using the
Geneva tracks along with the Lejeune atmospheres provides an even softer
radiation field in the $1-4$ Ryd energy range. In this case, the theoretical
grid falls below and to the left of the majority of the observed points 
on all three of the VO87 plots. This combination of tracks and atmospheres 
is therefore excluded with a good degree of certainty.

Finally, consider the PEGASE 2 models, which use the Padova tracks with the
Lejeune plus Clegg \& Middlemass PNN extended atmospheres. These models are 
characterised by the hardest radiation
field in the 1 - 4 Ryd region, and are the only ones that show the spectrum
becoming harder as the W-R stars turn on. This is a direct consequence of
the Clegg \& Middlemass PNN atmosphere models used for stars
with effective temperatures greater than 50000 K.  These are
the only set of models which encompass nearly all of the observed starbursts
on all three of the VO87 line diagnostic plots. Furthermore, individual
objects fall into similar regions of the theoretical grid in all three
diagnostic plots, indicating that there is some degree of consistency
achieved here. In essence, the ionization parameter appears to be limited to
the range $6\times 10^{6}\gtrsim q\gtrsim 6\times 10^{7}$ and the
metallicity range seems to cover most of the range; from about $0.2$ solar
up to nearly 3 times solar. However, the low-metallicity objects appear to
be rather rare in our sample, and most of the starbursts are consistent with
a metallicity 1 - 3 times solar.

We present the
grids of the continuous models for the 
[\ion{N}{2}]~$\lambda \lambda$6548,84 / [\ion{O}{2}]~$\lambda 
\lambda$3726,9~{\em vs. }[\ion{O}{3}]~$\lambda$5007 /H$\beta$ diagram and
the [\ion{N}{2}]~$\lambda \lambda$6548,84/[\ion{O}{2}]~$\lambda
 \lambda$3726,9~{\em vs. } [\ion{O}{3}]~$\lambda$5007/[\ion{O}{2}]~
$\lambda \lambda$3726,9 in Figures (12) and (12) for the use of the
astronomical community.

\subsection{Shock Excitation from Supernovae}

The starburst models presented here are based on pure photoionization
models, and therefore do not incorporate the effect of mechanical 
luminosity from supernovae shocks. The primary effect of the release 
of mechanical energy through shocks is to move the theoretical grids 
upwards, and lightly to the right on the VO87 line ratio diagnostic plots. 
The size of this effect depends on the relative importance of the shock and 
photoionization luminosity. 

The input mechanical energy luminosity ${\rm \stackrel{.}{E}_{mech}}$ 
produced by supernova events and winds is converted into optical line 
emission through 
radiative shocks. The importance of this line emission relative to that 
produced by photoionization depends on fraction of this 
energy, $\epsilon $,  which has been converted
to H$\beta $ flux compared to the H$\beta $ flux produced by recombinations in
the photoionized plasma:

\begin{equation}
\frac{L_{H\beta }(\mathrm{shock})}{L_{H\beta }(\mathrm{photo})}=\frac{
\epsilon {\rm \stackrel{.}{E}_{mech}}}{\alpha _{\mathrm{eff}}h\nu
_{H\beta }S_{*}}
\end{equation}

where $\alpha _{\mathrm{eff}}$ is the effective recombination coefficient of
hydrogen, and $S_{*}$ is the number of ionizing photons being produced by
the hot stars in the cluster. 

In the case of a supernova remnant, the shock becomes radiative, and the
Sedov-Taylor (adiabatic) expansion phase is terminated when its cooling
timescale, $\tau _{\mathrm{cool}}$ , becomes shorter than the dynamical
expansion timescale of the shell, $\tau _{\mathrm{\exp }}$. From standard
Sedov-Taylor theory, $\tau _{\mathrm{\exp }}=R/v_{\mathrm{s}}=5t/2$ where $R$
is the radius of the supernova remnant, $v_{\mathrm{s}}$ is the shock wave
velocity, and $t$ is the time since the supernova explosion. The radiative
shock-wave theory of Dopita \& Sutherland (1996) shows that the cooling
timescale, in years,  can be expressed as:

\begin{equation}
\tau _{\mathrm{cool}}\sim 200\frac{v_{100}^{4.4}}{Zn}
\end{equation}

where $v_{100}$ is the shock velocity in units of 100 km s$^{-1},$ $Z$ is
the metallicity of the plasma relative to solar, and $n$ is the pre-shock
density. We therefore find that, with $n\sim 350$ cm$^{-3}$, a supernova
remnant typically become radiative at a radius of 1 pc, when the shock
velocity is 600 km s$^{-1}.$ At this point, the supernova remnant is $\sim $%
600 years old, and its expansion timescale is $\sim $1500 yr. 

We therefore used a shock model with a velocity of 600 km s$^{-1}$ in a
spherical nebula of solar metallicity and $n\sim 350$ cm$^{-3}$ to calculate 
the possible contribution by supernova remnants to our photoionization models.
We assumed a total mechanical luminosity in SNR of 
$6\times 10^{41}\,{\rm erg/s/M_{\odot}/yr}$ \citep{Leitherer99}.  We
calculated a star formation rate of $\sim 3.4 \,{\rm M_{\odot}/yr}$
from the average \LIR for the starbursts in our sample,
following the prescription given in \citet{Kennicutt98} (note that \LIR 
is defined as \LFIR in \citet{Kennicutt98}).  Assuming the IR 
luminosity is distributed evenly throughout the galaxy and a minimum
size of 7 kpc, the SFR reduces to $\sim 0.07\,{\rm M_{\odot}/yr}$ in 
the observed aperture.
We note that this is still higher than that found using the \Ha~luminosity 
of our template starburst (within the 1 kpc aperture) of 
$\sim 0.04~{\rm M_{\odot}/yr}$.  This is to be expected as dust 
absorption will reduce the \Ha~derived SFR compared with that derived
using the FIR luminosity.  We computed a 600 km/s shock model and a
spherical ionized precursor with a shocked spherical radius of 1 pc.  
This model had a mechanical luminosity of $3.6 \times 10^{39}$ erg/s.  
As we expect the total mechanical luminosity to 
be $\sim 4 \times 10^{40}$ erg/s, we expect to obtain on average
11.2 SNR within our 1 kpc aperture at any one time.  
The luminosity produced in the shock and precursor are given in Table (2).

From this SNR model, the contribution to \Hb~emission in the
starbursts due to SNRs would be about 16-20\%.  The contribution to the 
\Hb~flux from starbursts is determined more or less directly by the
star formation history and the IMF.  The contribution to \Hb~from 
SNRs is determined by the velocity and the total area covered by 
the SNR shocks, which in turn 
depend on the number of SNRs (determined by the SFR and IMF) and weakly 
on the density.  At low densities, SNRs are larger and have lower velocities
than at higher densities, and have similar overall luminosities in \Hb.

The \OIII\ emission from the starbursts, on the other hand, is not a 
function of the SFR, rather it is a measure of the ionization parameter $q$ 
of the radiation field within the starburst which is a function of the 
density and mass distribution (ie filaments vs uniform) of the gas.  Thus the 
\OIII~in the starburst models varies from bright to negligible 
levels compared to \Hb.  The \OIII ~mission from the SNRs 
is constrained by the chosen model geometry and is therefore 
determined by the SFR and the average density, not by the global
ionization parameter. The ionization parameter for each SNR is 
determined internally by the SNR model without reference to the global 
starburst value.

The observed total \OIIIHb~ratio is the sum of the starburst and SNR
contributions; 

\begin{equation}
\log([{\rm OIII}]/{\rm H}\beta)  = \log([{\rm OIII}]_{\rm starb} + 
[{\rm OIII}]_{\rm SNR})
- \log({\rm H\beta_{starb}+H\beta_{SNR}}).
\end{equation}

Thus, in the limit of $[{\rm OIII}]_{\rm starb}=0$ due to low global $q$ in 
the starbursts, the observed \OIIIHb~ratio will reach a lower limit of 

\begin{equation}
\log([{\rm OIII}]_{\rm SNR}) - \log({\rm H\beta_{starb}+H\beta_{SNR}})
\ge 0.0 
\end{equation}

with a density of 350 cm$^{-3}$ and the SFR used here ($\sim 0.07\,{\rm M_{\odot}/yr}$ in 
the observed aperture of 1 square kpc).

This lower limit is not observed, and the actual lower limit is $\log({\rm
[OIII]/H}\beta) \approx -1.0$, a factor of 10 less than the SNR model limit.
Clearly the strong $[{\rm OIII}]_{\rm SNR}$ contribution modeled here is not
compatible with the observations.  This incompatibility may be due to either
the number of SNR we derived, or that the average density in the SNR
environment is less than 350 cm$^{-3}$.  We conclude that the SNR
contribution to the \OIIIHb~ratio is $\ll20$\% and is probably a factor of 10
less ($\sim 2$\%),  and can be neglected for the starburst models derived
here.   We are currently investigating this SNR contribution  to starbursts
further by considering a range of densities and lower shock velocities to
determine the level at which the SNRs model is compatible  with the
observations.  We expect to find that velocities in the 200-300 km/s  range
will be more compatible with observations, resulting in a small 
$[{\rm OIII}]_{\rm SNR}$
contribution, negligible in all but the starbursts with the  lowest ionization
parameter.

\section{Wolf-Rayet Emission in Starburst Galaxies}

Wolf-Rayet features were first found in the dwarf emission galaxy
He2-10 \citep{Allen76}.  As more such features were discovered in 
galaxies, \citet{Osterbrock82} defined Wolf-Rayet (W-R) galaxies as those
galaxies which contain broad stellar emission lines in their spectra and
therefore contain large numbers of W-R stars. The large numbers of W-R stars
are thought to be a result of present or very recent star formation. 
\citet{Kunth85} searched for W-R emission in a sample of blue emission-line galaxies.  They found a positive detection in one galaxy and suspected 
in 14 others and suggested that W-R stars are preferentially detected in
low redshift galaxies.  The first catalogue of W-R galaxies 
was presented by \citet{Conti91}, showing that
W-R galaxies can be easily distinguished by their broad $\lambda$4686 [
\ion{He}{2}] emission feature, or in some cases a broad line at 4640\AA~due 
to \ion{N}{3}. High resolution long-slit observations of a sample
of Wolf-Rayet galaxies were carried out recently by \citet{Guseva00}. Nearly
all the galaxies in their sample show broad W-R emission consisting of an
unresolved blend of \NIII~$\lambda$4640, \CIII~$\lambda$4650, 
\FeIII~$\lambda$4658, and \HeII~$\lambda$4686 emission lines.
They also found weaker W-R emission lines \NIII~$\lambda$4512 and 
\SiIII~$\lambda$4565 in some galaxies.

The signal-to-noise ratios of our spectra are not sufficient in individual
galaxies to detect these signatures of W-R emission. We have therefore 
constructed a template ``average'' warm infrared starburst spectrum from 
the 56 starburst galaxies in our sample which have SNRs at H$\beta $ of 60 
or greater and for which the zero-redshift blue wavelength cut-off is 
lower than $\lambda $4620.  We note that this average spectrum is not 
representative of our sample of warm infrared starbursts because 
selecting for high SNR galaxies may bias the average spectrum towards 
young and hence more luminous starbursts. However, it is useful simply to
assist in the search for W-R features. The average spectrum is shown
in Figure (14) and in detail in Figure (15). The positions of the expected 
W-R features are marked on Figure (15).

The mean age of the average warm infrared starburst can be estimated
from the H$\beta $ absorption equivalent width, bearing in mind that 
dust absorption may also contribute to the absorption line profile and
has not been taken into account.  The H$\beta $ absorption equivalent
width was found by simultaneously fitting the H$\beta $ absorption and
emission lines with gaussians in the IRAF task {\it ngaussfits}. We
find that the H$\beta $ absorption equivalent width is $\sim 3.6$~\AA which
corresponds to an upper limit age of $\sim 7$ Myr for a continous star 
formation model at solar metallicity \citep{Gonzalez99}. However, this
is probably an underestimate, since the H$\beta $ absorption equivalent 
width will be reduced by any underlying old star
population. Nonetheless, there is likely to be likely a selection affect
towards more youthful starbursts, in that the bright starbursts tend to be 
both younger and intrinsically more luminous.

In the ``average'' spectrum of Figure (15), there
appears to be marginal detections of \FeIII~$\lambda $4658, and 
\HeII~$\lambda $4686 at the 2$\sigma $ level. These lines do not
appear broad, however this may simply be an effect of the low SNR for these
lines.

The \HeII~$\lambda $4686 emission
in the majority of our galaxies provides an important constraint on the 
Schmutz extended atmospheric modeling.  The 
\logHeIIHb ratio for our template starburst galaxy 
is $\sim -1.6$.  Both the MAPPINGS~III models with PEGASE and 
STARBURST99 (Lejeune atmospheres) stellar ionizing continua produce 
\logHeIIHb~around $-6$, while the MAPPINGS~III models 
with STARBURST99 (Lejeune + Schmutz atmospheres) stellar ionizing continuum
produces \logHeIIHb around $-1.7$, consistent
with that observed with our template starburst galaxy.  We note
again that our template starburst galaxy has been composed of the
starbursts in our sample with the highest SNRs, and so may be biased towards
the most luminous and therefore the youngest starbursts in our sample.
This \HeIIHb ratio for our template starburst
may therefore be biased towards those with more W-R stars than the
average starburst galaxy in our sample.  Clearly we require a hard EUV 
field in the 1-4
Ryd regime to model our starburst galaxies on the optical diagnostic
diagrams such as that provided with the PEGASE 2 model.  However, as
we would expect the Schmutz extended atmospheric modeling to be more 
appropriate to the modeling of our starburst galaxies, we believe
that the inclusion of continuum metal opacity (or continuum metal blanketing) 
in the stellar population synthesis models using the Schmutz extended 
atmospheres is a possible solution.  

\section{Continuum Metal Opacity}

For many years, optical depth estimates in the ISM were found using the 
hydrogen 21cm
line and the assumption of a uniform gas density.  As a result of the low
optical depths obtained, it was thought that the ISM should be opaque to 
radiation in the EUV \citep{Aller59}.  The discovery that the ISM is 
inhomogeneous overturned this conclusion and \citet{Cruddace74} established 
an estimate of the effective absorption cross-section of the ISM at EUV
wavelengths using determinations of cross-sections and abundances.  
\citet{Cruddace74} showed that some EUV radiation should be 
able to be observed over considerable distances. More recent determinations 
of cross-sections and abundances have been used by \citet{Rumph94} to 
provide a new estimate of the effective absorption cross-section of the 
ISM at EUV wavelengths.  

The relative difficulty in obtaining EUV spectra to compare with theoretical
estimates of opacity in the EUV continuum described above has decreased 
significantly over the last decade with the aid of 
EUV telescopes such as EUVE (Extreme Ultraviolet Explorer), ROSAT Wide Field 
Camera, ALEXIS (Array of Low Energy X-ray Imaging Sensors), 
FUSE (Far Ultraviolet Spectroscopic Explorer), 
and the Extreme-Ultraviolet Imaging Telescope (EIT). These telescopes have
allowed observations of many stellar objects which have greatly advanced
the modeling of stellar atmospheres \citep[eg.]{Cassinelli95} and have 
allowed the EUV study of some Seyfert galaxies (eg. \citet{Hwang97}).  
However the EUV spectrum in starburst galaxies continues to remain unseen as 
a result of the weakness of the EUV continuum in starburst galaxies due to 
absorption.  It is therefore necessary to rely on 
theoretical predictions of the EUV spectrum from 
stellar population synthesis codes.

Although the EUV spectrum must be estimated using theoretical modeling, the 
reprocessing of the EUV spectrum into optical emission lines allows
limits to be placed on the shape of the EUV spectrum in starburst galaxies.  
As concluded in previous sections, we require a harder EUV field between 
the 1-4 Ryd regime than is provided by the stellar population synthesis 
models with Schmutz extended atmospheres.  Currently, only continuous 
opacities due to helium are included in the models, as continuous metal 
opacities are relatively unimportant for many studies.  We believe that the 
inclusion of continuum metal opacities in these models would provide a
suitable shape in the EUV, enabling the models to reproduce the starburst
sequence on the standard optical diagnostic diagrams.

Continuum metal opacities, as with hydrogen and helium opacities,
are a result of bound-free transitions, ie photoionization of the
metals.  Continuum metal opacity would allow some fraction of the 
radiation with energies greater than 4 Ryd to be absorbed and re-emitted 
less than 4 Ryd.  The fraction of radiation absorbed by metals depends
on their individual absorption cross-sections and abundances.
The resulting EUV continuum would have a softer 
continuum above the \HeII~ionization limit, as a result of carbon
opacities.  The spectrum would also have a harder but fainter EUV field 
between the 1-4 Ryd 
regime, as required by the position of our starburst galaxies on the
optical diagnostic diagrams.  Note that if this were the case, we might expect that 
the models with Schmutz extended atmospheres would produce too much flux in 
the \HeII~$\lambda 4686$ line compared with our observations, contrary to
our findings.   We conclude that while continuum
metal blanketing may be a possible solution to the discrepancy seen 
between our observations and the models using Schmutz extended atmospheres,
it may not be the only solution.

\section{Extreme Starburst Classification Line}

In order to place a theoretical upper limit for starburst models
on the optical diagnostic diagrams, we used the PEGASE 2.0 grids, since 
these provide the hardest EUV spectrum, and therefore give a theoretical
grid which is placed both highest and furthest to the right on the VO87 line
ratio diagnostic diagrams. 

We have shown that, with a realistic range of metallicities ($Z=0.1-3.0$) 
and ionization parameter $q$ (cm/s) 
in the range $5\times 10^{6} \le q \le
3\times 10^{8}$ (or $-3.5 \le \log {\cal U} \le -2.0$),
continuous starburst models produced by any modeling procedure always 
fall below and to the left of an empirical limit on the 
\NII/${\rm H \alpha}$ {\it vs.} \OIII/${\rm H \beta}$,
\SII/${\rm H \alpha}$ {\it vs.} \OIII/${\rm H \beta}$ and 
\OI/${\rm H \alpha}$ {\it vs.} \OIII/${\rm H \beta}$ diagrams. This is
due to the two-parameter grid of the ionization parameter and metallicity 
folding back upon itself.  Therefore, no 
combination of these parameters can generate a theoretical point above this
fold.  The lines dividing the theoretical starburst region from objects of
other types of excitation are shown in Figure (16). These can be 
parametrized by the following simple fitting formulae, which have the 
shape of rectangular hyperbolae;

\begin{equation}
\log \left(\frac{{\rm [OIII]\,\lambda 5007}}{{\rm H}\beta}\right) = \frac{0.61}
{\log ({\rm [NII]}/{\rm H}\alpha) -0.47}+1.19
\end{equation}

\begin{equation}
\log \left(\frac{\rm [OIII]\,\lambda 5007}{{\rm H}\beta}\right) = \frac{0.72}
{\log ({\rm [SII]\,\lambda \lambda 6717,31}/{\rm H}\alpha) -0.32}
+1.30
\end{equation}

\begin{equation}
\log \left(\frac{\rm [OIII]\,\lambda 5007}{{\rm H}\beta}\right) = \frac{0.73}
{\log ({\rm [OI]\,\lambda 6300}/{\rm H}\alpha) +0.59}+1.33
\end{equation}

The shape and position of this maximum starburst line has not been 
previously established from theoretical models, since the shape of the 
ionizing spectrum of the cluster has not been known to sufficient accuracy. 
VO87 have attempted to establish both the position and the shape of this 
boundary in a semi-empirical way, using both observational data from 
the literature and a combination of models available to them at that time.

The theoretical boundaries for starbursts defined by 
equations (1)-(3) provide us for the first time with a theoretical (as 
opposed to a semi-empirical boundary) for the region occupied by
starbursts in these diagnostic plots. In view of the potential for errors 
in the modeling which may flow from errors in the assumptions made 
in the chemical abundances, chemical depletion factors, the slope of the 
initial mass function, and the evolutionary tracks and the stellar 
atmosphere models used, we have indicated a ``best guess'' estimate of 
these errors as dashed lines in Figure (16).

These theoretical upper limits for starburst galaxies have been used
in \citet{Kewley00a} to classify the galaxies in our sample along with
an 'extreme' mixing line produced using our shock modeling to classify
galaxies into starburst, LINER and AGN types.  The strength of our
theoretical starburst classification line can be seen by observing the
number of galaxies which have `ambiguous' classifications.  These
galaxies are those which fall within the starburst region of one or
two of the diagnostic diagram(s) and the AGN region of the remaining
diagram(s). We found that only 6\% of our sample have ambiguous 
classifications using our theoretical extreme starburst line, 
compared with 16\% ambiguous classifications using the standard VO87 
classification lines.  These results indicate that our theoretical 
starburst line 
is a reliable tool for optically classifying galaxies into starburst 
and AGN types, and is more consistent from diagram to diagram than the
conventional VO87 method.

\section{Conclusions}

We have presented a comparison between the stellar population synthesis
models PEGASE 2 and STARBURST99 using a large sample of 157 warm infrared
starburst galaxies. The main differences between the two synthesis models
are the stellar tracks and stellar atmosphere prescriptions. 
PEGASE and STARBURST99 were used to generate the
spectral energy distribution (SED) of the young star clusters. MAPPINGS~III
was used to compute photoionization models which include a self-consistent
treatment of dust physics and chemical depletion. The standard optical
diagnostic diagrams are indicators of the hardness of the EUV radiation
field in the starburst galaxies. These diagnostic diagrams are most sensitive to the
spectral index of the ionizing radiation field in the 1-4 Rydberg region. We
find that warm infrared starburst galaxies contain a relatively hard EUV
field in this region. The PEGASE ionizing stellar continuum is harder in the
1-4 Ryd range than that of STARBURST99, most likely due to the different
stellar atmospheres used for Wolf-Rayet stars.  

We have constructed an average spectrum of the high SNR warm infrared
starbursts in our sample in order to look for Wolf-Rayet signatures. We find
detections of \ion{C}{4} $\lambda$4658, and \ion{He}{2} $\lambda$4686 at
the 2$\sigma$ level, indicating W-R activity, and constraining the
Schmutz extended
atmospheric modeling.  We require a hard EUV field in the 1-4
Ryd regime to model our starburst galaxies on the optical diagnostic
diagrams such as that provided with the PEGASE 2 model.  However, as
we would expect the Schmutz extended atmospheric modeling to be more 
appropriate to the modeling of our starburst galaxies, we believe
that one solution would be to include continuum metal blanketing in 
the stellar population synthesis models using the Schmutz extended 
atmospheres.  Continuum metal blanketing would allow much of the 
radiation with energies greater than 4 Ryd to be absorbed and 
re-emitted at energies less than 4 Ryd.  The resulting
EUV continuum would have a softer continuum above the \HeII~ 
ionization limit and a harder EUV field between the 1-4 Ryd 
regime, as required by the position of our starburst galaxies on the
optical diagnostic diagrams.  SNRs could also contribute to the hardness
of the EUV field, although our models and observations suggest that this is 
likely to be $\ll20$\%, insufficient to cause the discrepancy 
between our starburst galaxies and the models using Schmutz extended 
atmospheres.

We use the starburst grids produced with the PEGASE EUV
ionizing radiation field and our MAPPINGS~III models to parametrize an
extreme starburst line which is useful in classifying galaxies
into starburst or AGN types.  In a previous paper \citep{Kewley00a}, 
we showed that this theoretical classification scheme produces reliable
classifications with less ambiguity than the classical
\citet{Veilleux87} empirical method.

\section{Acknowledgements}

We thank Claus Leitherer and Brigitte Rocca-Volmerange for 
helpful discussions and for the use of STARBURST99 and PEGASE 2. 
We thank the referee for his constructive comments which 
helped this to be a better paper.  We also thank the staff at Siding 
Springs Observatories for their assistance during our spectroscopy 
observations.  L. Kewley gratefully acknowledges support from the 
Australian Academy of Science Young Researcher Scheme and the French 
Service Culturel \& Scientifique.

\clearpage

\newpage
\epsscale{0.75}
\plotone{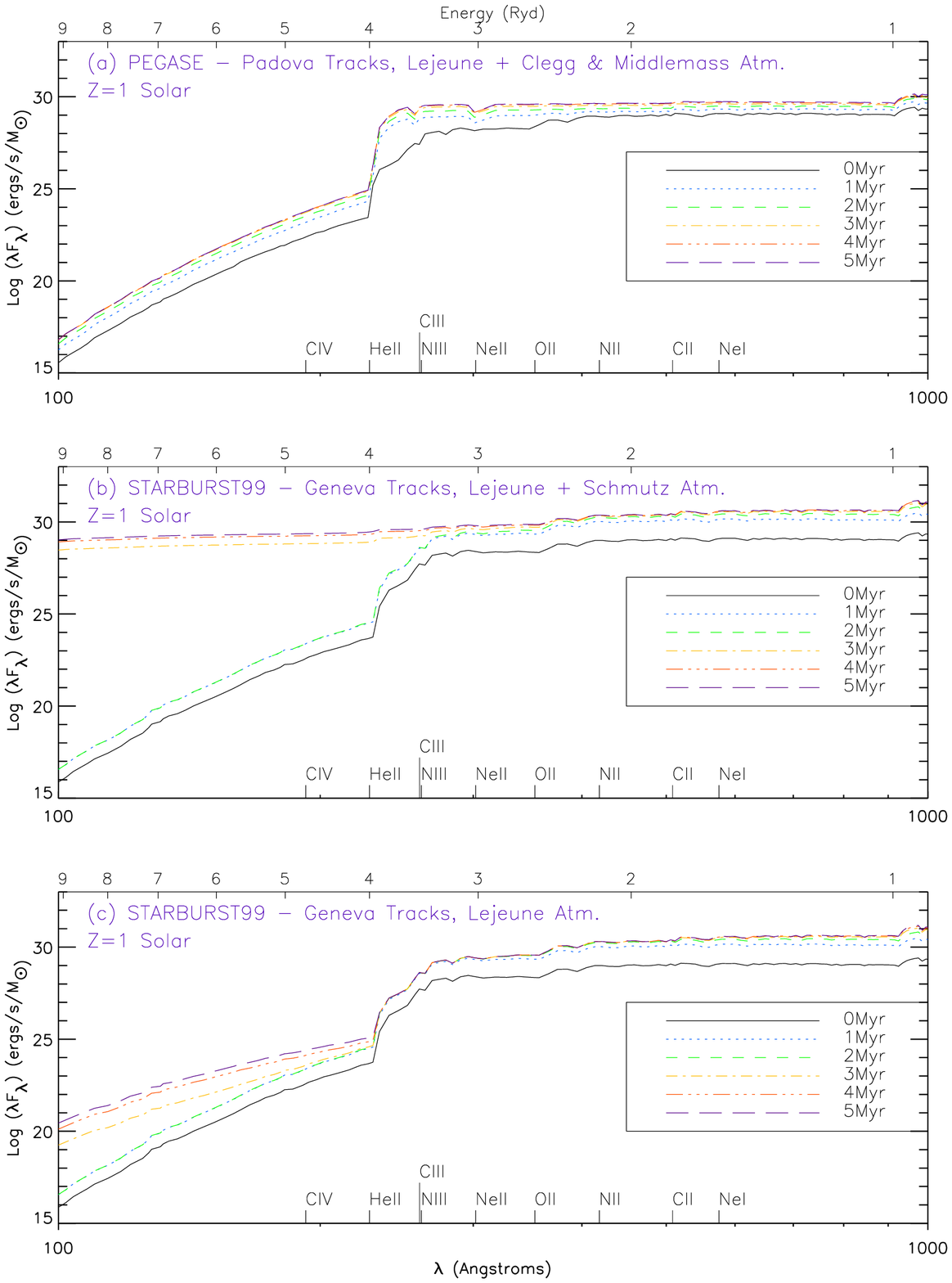}
\begin{small}
\figcaption[figure1.ps]{The ionizing spectral energy distribution (normalised to
the flux at the Lyman limit) as a function of age of the starburst for
(a)the PEGASE models, (b)  and the Lejeune plus Schmutz atmospheres
and (c) the STARBURST99 models for Lejeune atmospheres.  Ionization edges 
are shown for some elements. Note the rapid intitial evolution as the 
Wolf-Rayet stars
evolve, and the convergence toward an asymptotic solution when a stochastic
balance between star births and star deaths is achieved for all masses which
contribute to the EUV continuum. All models produce very similar results at
zero age, but differ markedly at later times, mainly reflecting differences
between the extended stellar atmosphere models used (\citet{Clegg87} for PEGASE
and \citet{Schmutz92} for STARBURST99).    \label{fig1}}
\end{small}

\newpage
\epsscale{0.85}
\plotone{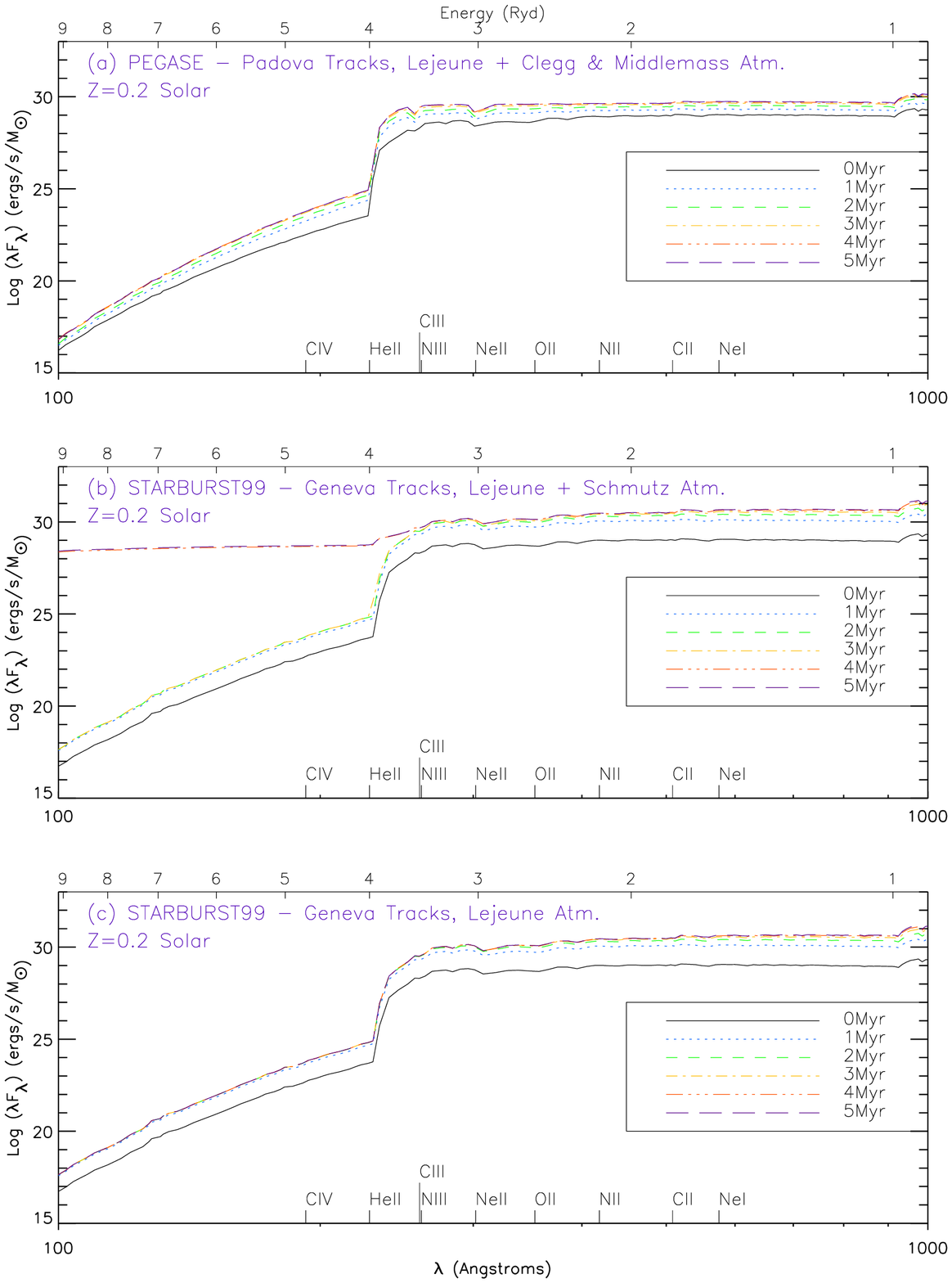}
\figcaption[figure2.ps]{As in Figure 1, except Z=0.2 $\times$ solar metallicity. 
{\label{fig2}}}

\newpage
\plotone{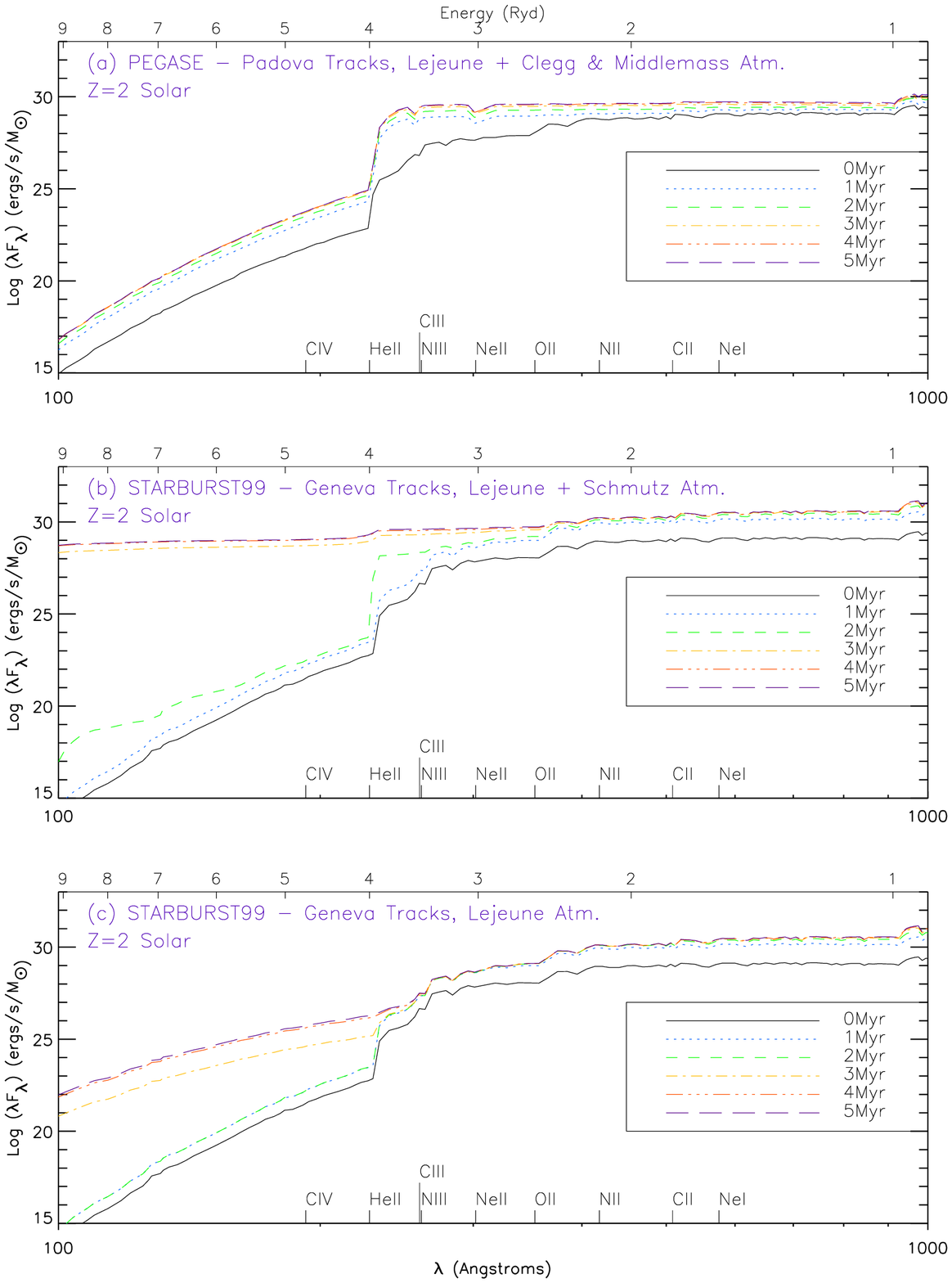}
\figcaption[figure3.ps]{As in Figure 1, except Z=2 $\times$ solar metallicity. 
{\label{fig3}} }

\newpage
\plotone{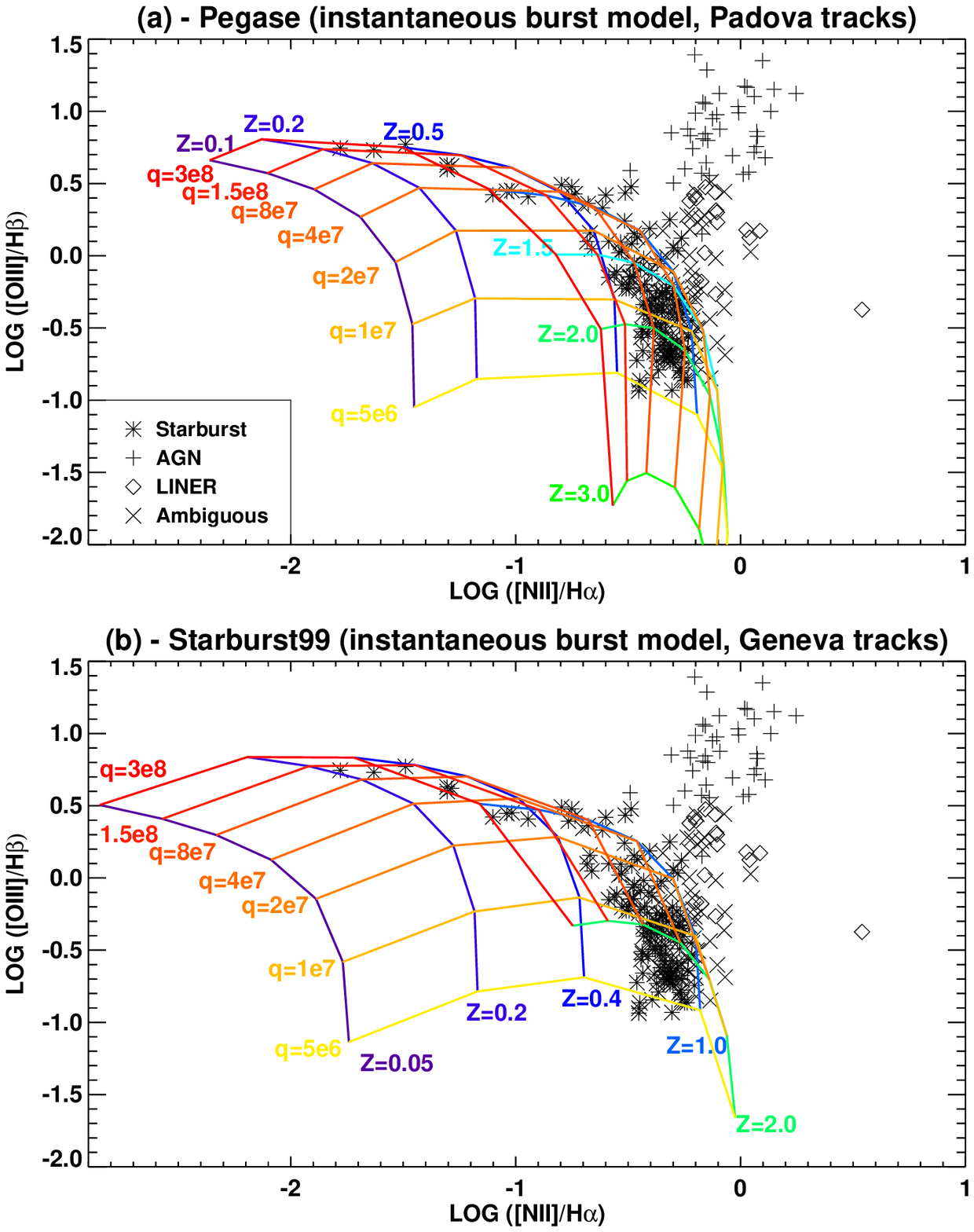}
\figcaption[figure4.ps]{The \cite{Veilleux87} diagnostic diagram $\log \mathrm{
[NII]/H}\alpha \log \emph{vs}$ $\log \mathrm{[OIII]/H}\beta $ for (a) the
instantaneous zero-age starburst models based on the PEGASE spectral energy
distribution, (b) instantaneous zero-age starburst models based on the
STARBURST99 spectral energy distribution with Lejeune plus Schmutz
atmospheres. The theoretical grids of ionization parameter and chemical
abundance are shown in each case. {\label{fig4}} }

\newpage
\plotone{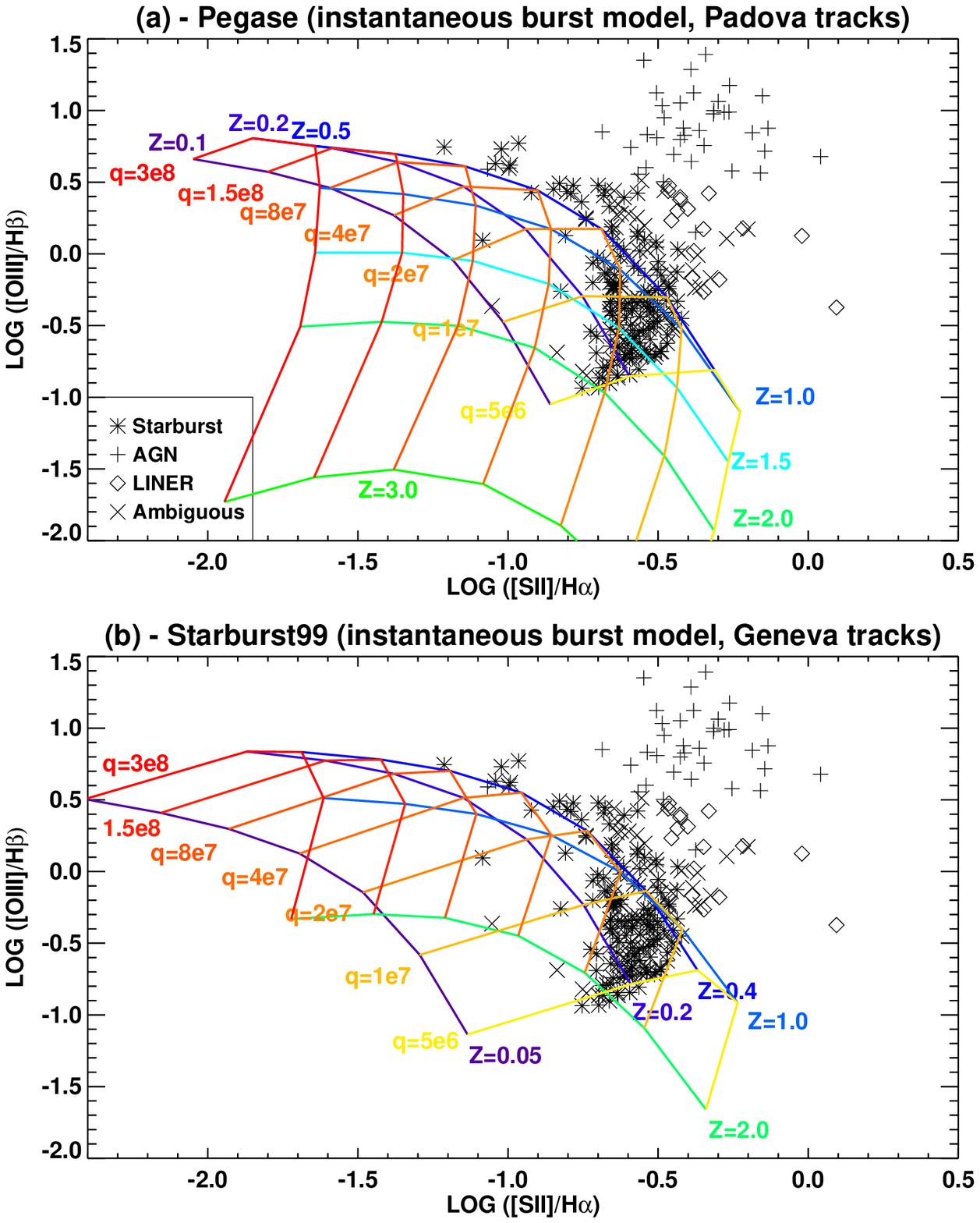}
\figcaption[figure5.ps]{As Figure (4) but for the diagram $\log \mathrm{\ [SII]/H}
\alpha $ $\emph{vs}$ $\log \mathrm{[OIII]/H}\beta$.{\label{fig5}} }

\newpage
\epsscale{0.90}
\plotone{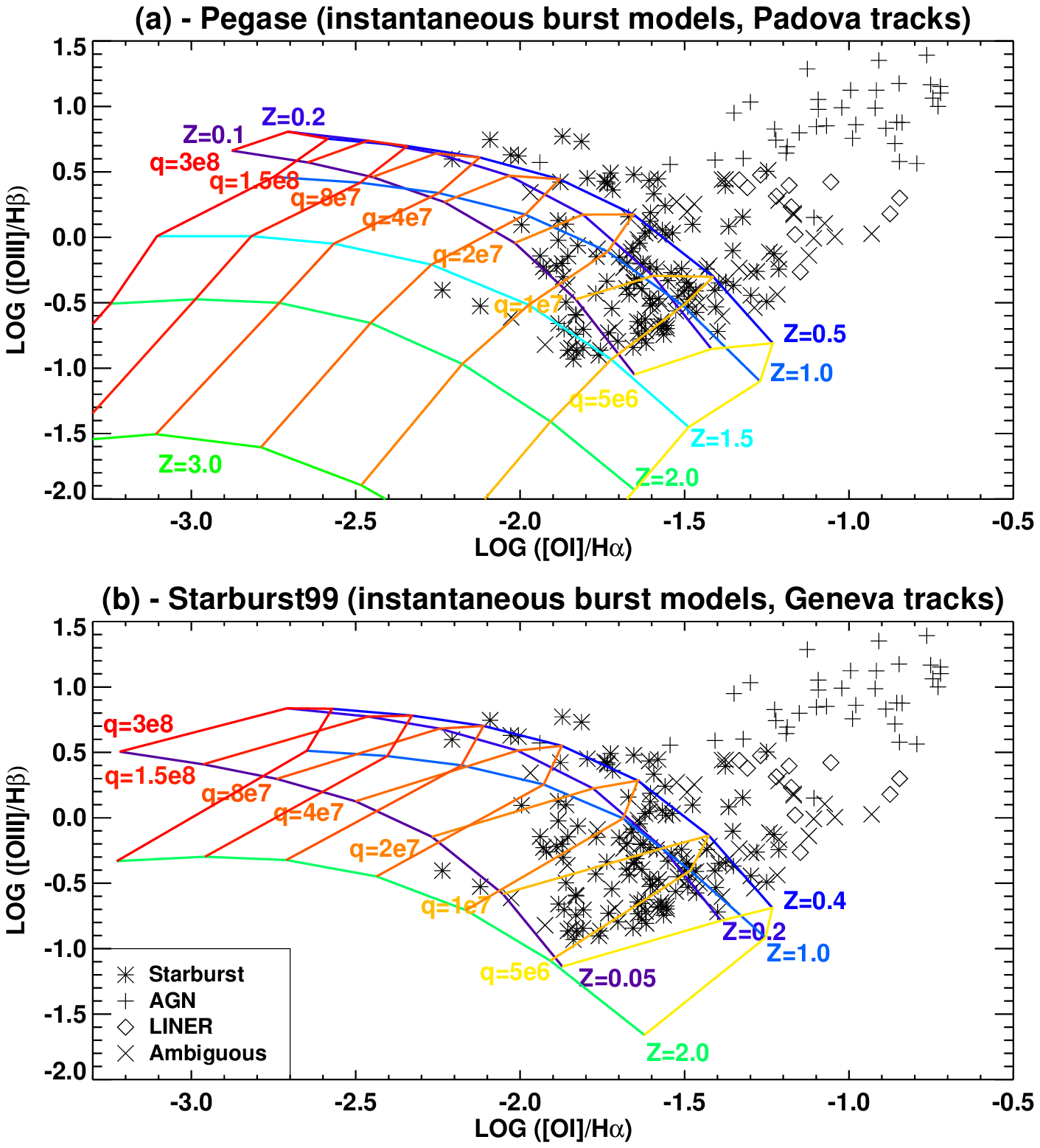}
\figcaption[figure6.ps]{As Figure (4) but for the diagram $\log \mathrm{[OI]/H }
\alpha $ $\emph{vs}$ $\log \mathrm{[OIII]/H}\beta .$
{\label{fig6}} }

\newpage
\epsscale{0.85}
\plotone{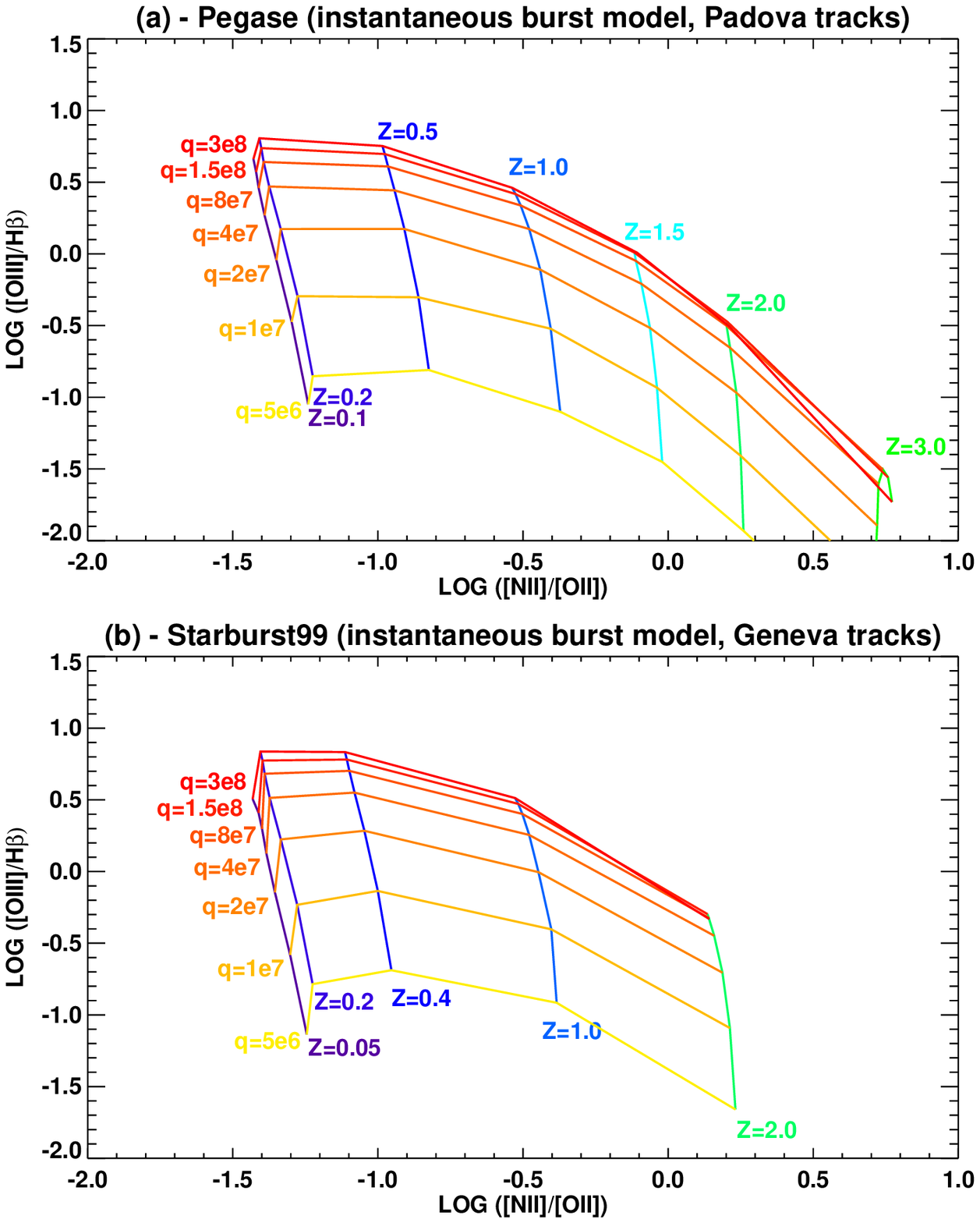}
\figcaption[figure7.ps]{As Figure (4) but for the diagram $\log \mathrm{[NII]/[OII]} $ 
$\emph{vs}$ $\log \mathrm{[OIII]/H}\beta .$  We do not have [OII] observations
for the galaxies in our sample, but we provide this diagram for the 
use of the astronomical community. {\label{fig7}} }

\newpage
\plotone{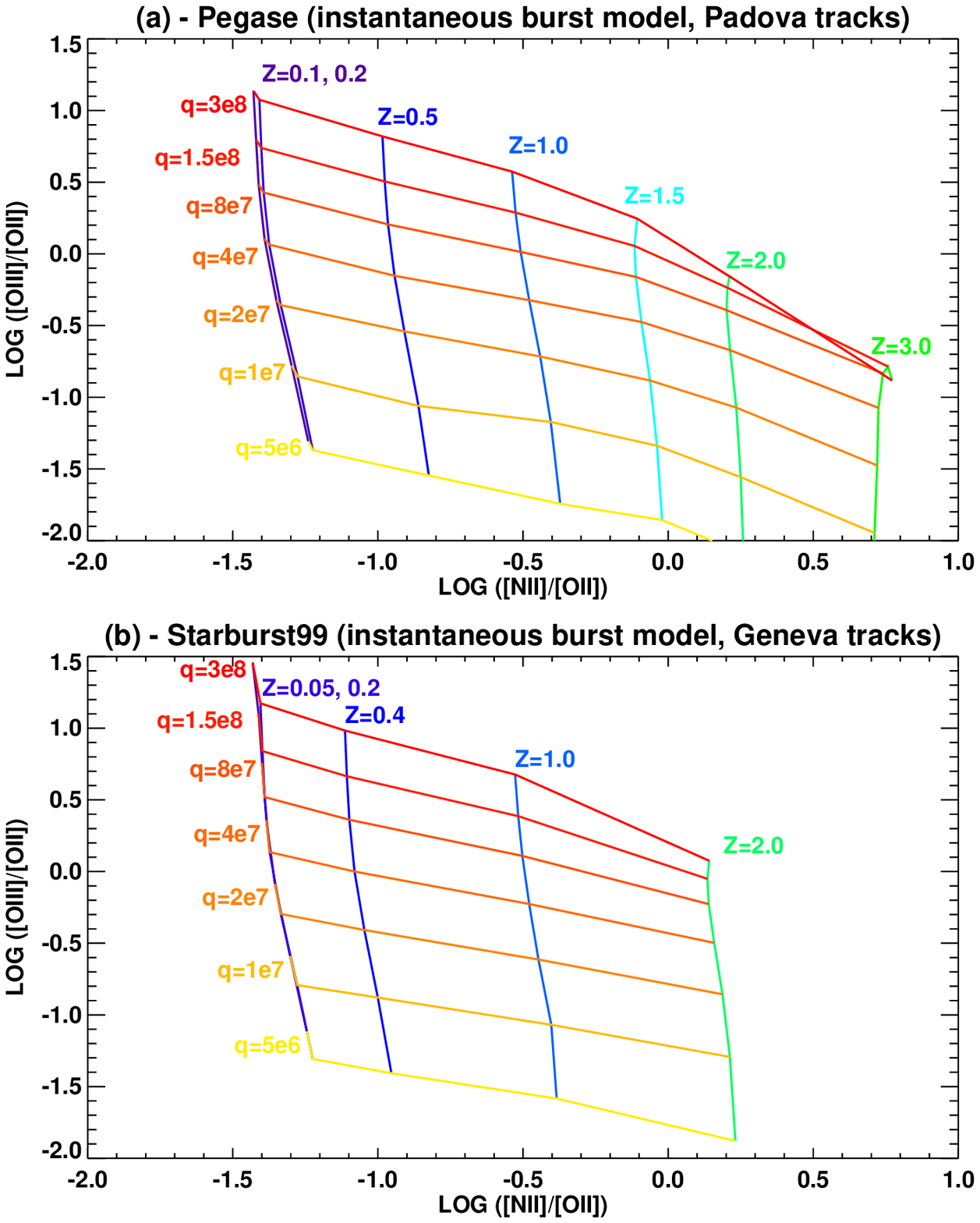}
\figcaption[figure8.ps]{As Figure (4) but for the diagram $\log \mathrm{[NII]/[OII]} $ 
$\emph{vs}$ $\log \mathrm{[OIII]/[OII]}$ We do not have [OII] observations
for the galaxies in our sample, but we provide this diagram for the 
use of the astronomical community. {\label{fig8}} }

\newpage
\epsscale{0.75}
\plotone{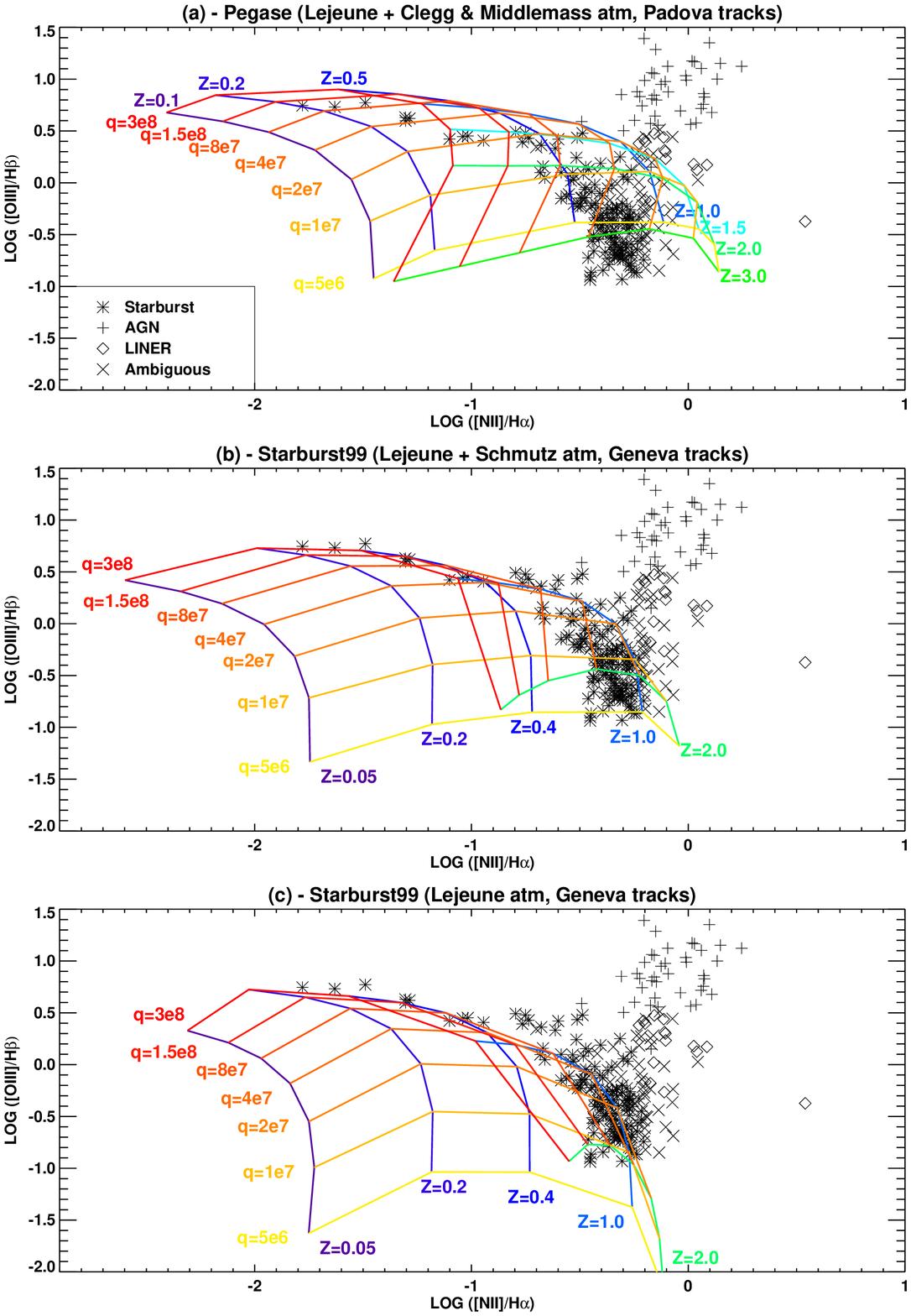}
\figcaption[figure9.ps]{The \cite{Veilleux87} diagnostic plot $\log \mathrm{
[NII]/H}\alpha \log \emph{vs}$ $\log \mathrm{[OIII]/H}\beta $ for (a) the
continuous starburst models based on the PEGASE spectral energy
distribution, (b) continuous starburst models based on the STARBURST99
spectral energy distribution with Lejeune plus Schmutz atmospheres, and (c)
continuous starburst models based on the STARBURST99 spectral energy
distribution with Lejeune atmospheres. The theoretical grids of ionization
parameter and chemical abundance are shown in each case. 
{\label{fig9}} }

\newpage
\epsscale{0.80}
\plotone{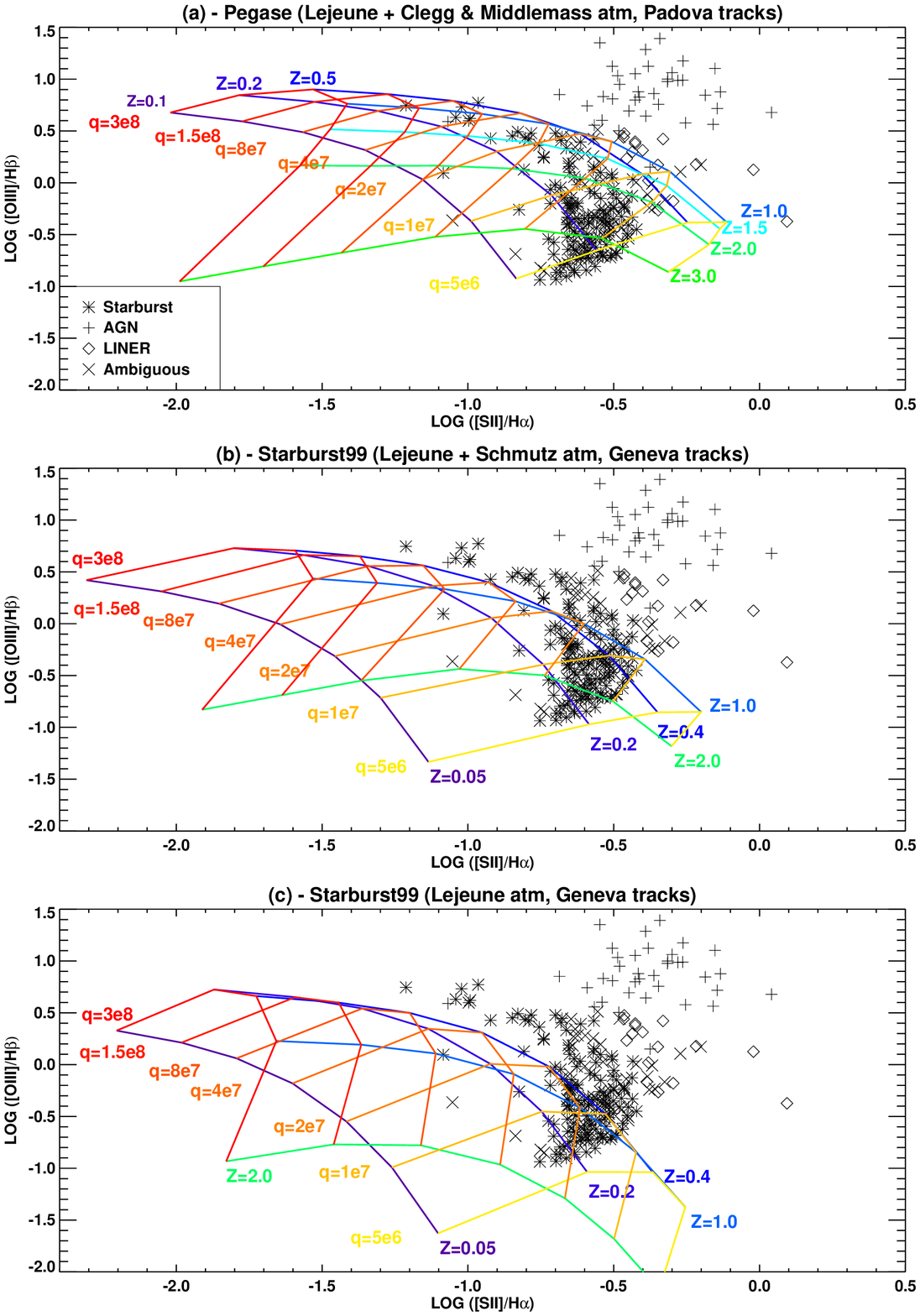}
\figcaption[figure10.ps]{As Figure (9) but for the plot $\log \mathrm{\ [SII]/H}
\alpha $ $\emph{vs}$ $\log \mathrm{[OIII]/H}\beta .$
{\label{fig10}} }

\newpage
\plotone{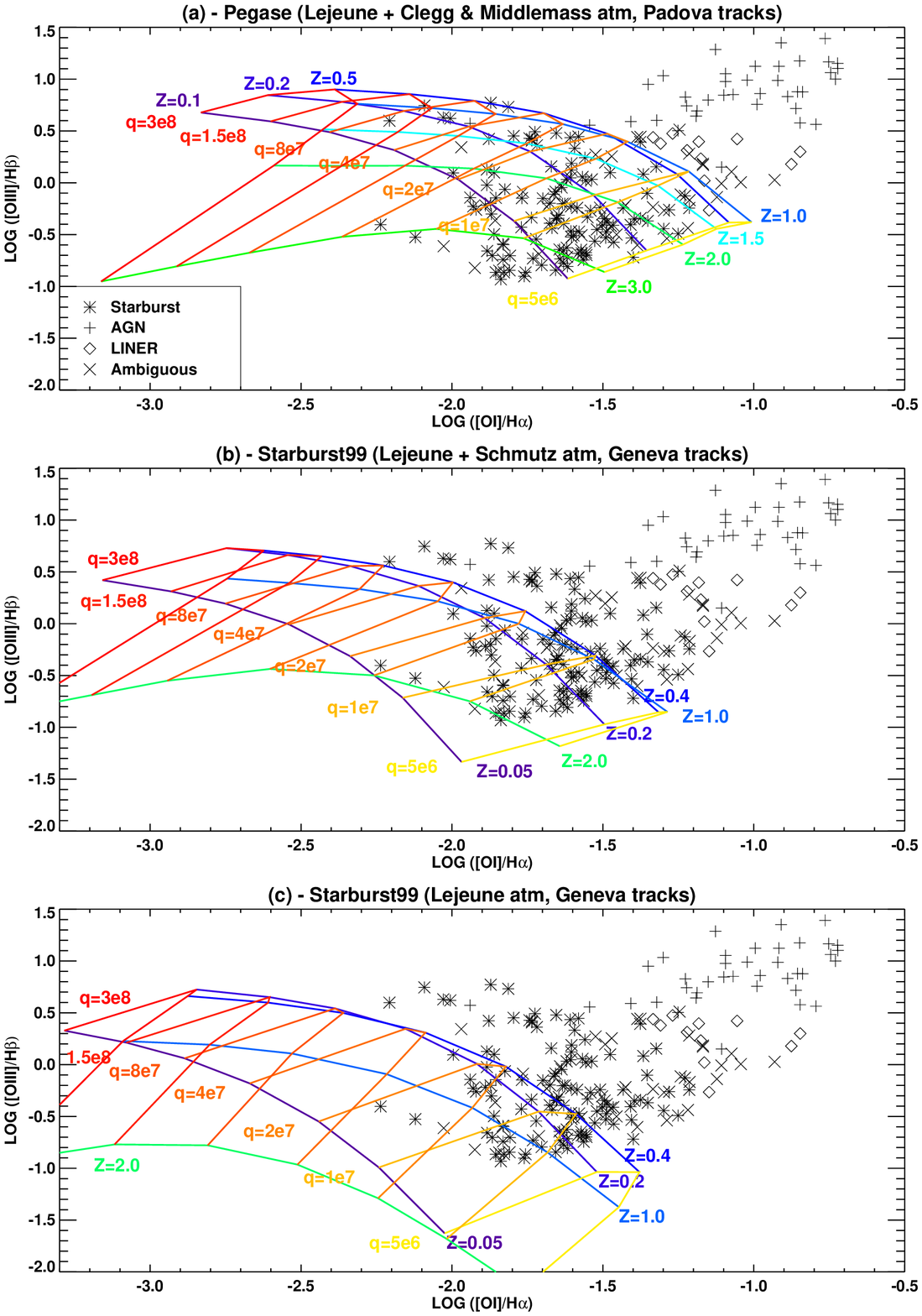}
\figcaption[figure11.ps]{As Figure (9) but for the plot $\log \mathrm{[OI]/H }
\alpha $ $\emph{vs}$ $\log \mathrm{[OIII]/H}\beta .$
{\label{fig11}} }

\newpage
\plotone{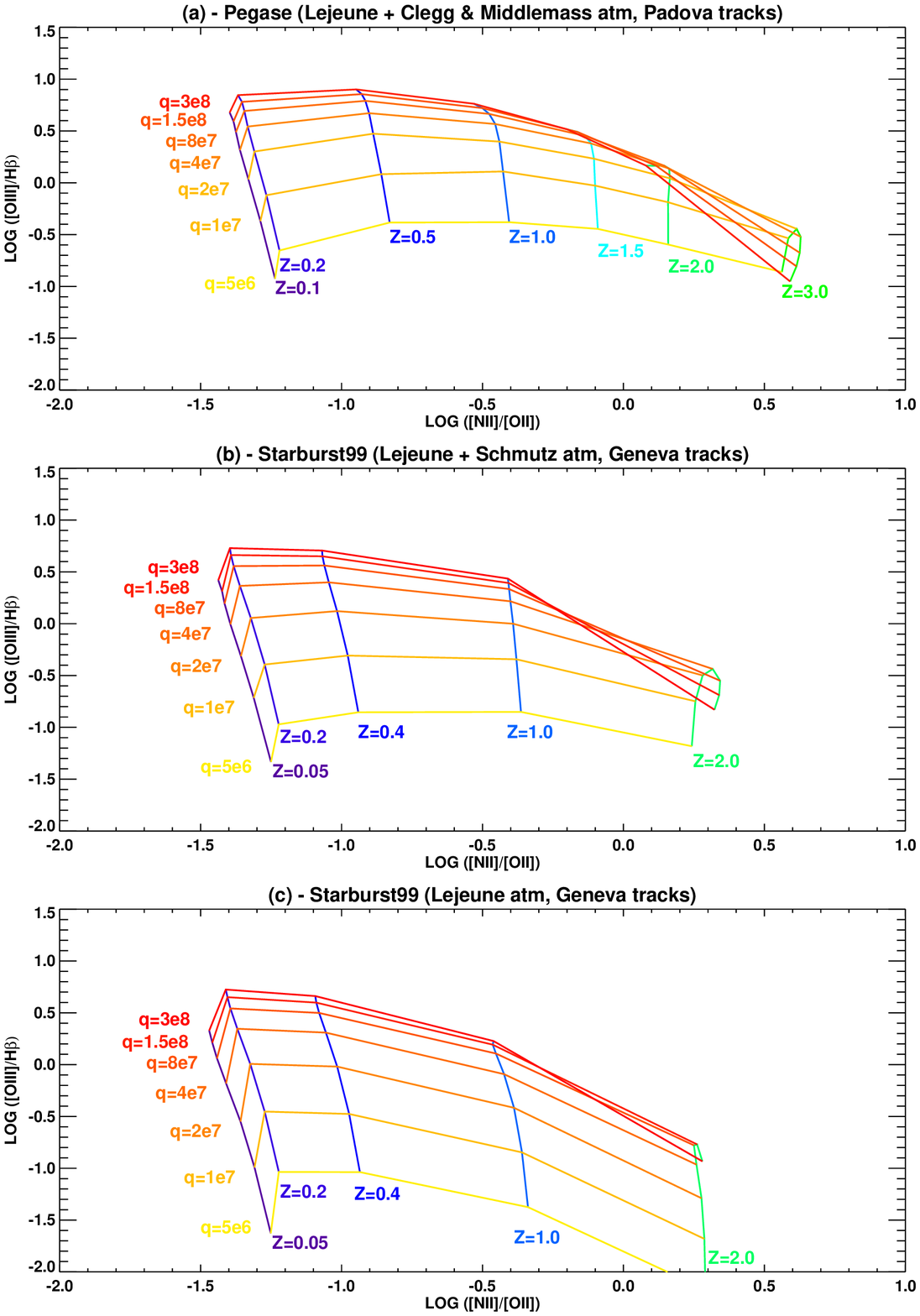}
\figcaption[figure12.ps]{As Figure (9) but for the diagram $\log \mathrm{[NII]/[OII]} $ 
$\emph{vs}$ $\log \mathrm{[OIII]/H}\beta .$  We do not have [OII] observations
for the galaxies in our sample, but we provide this diagram for the 
use of the astronomical community. {\label{fig12}} }

\newpage
\plotone{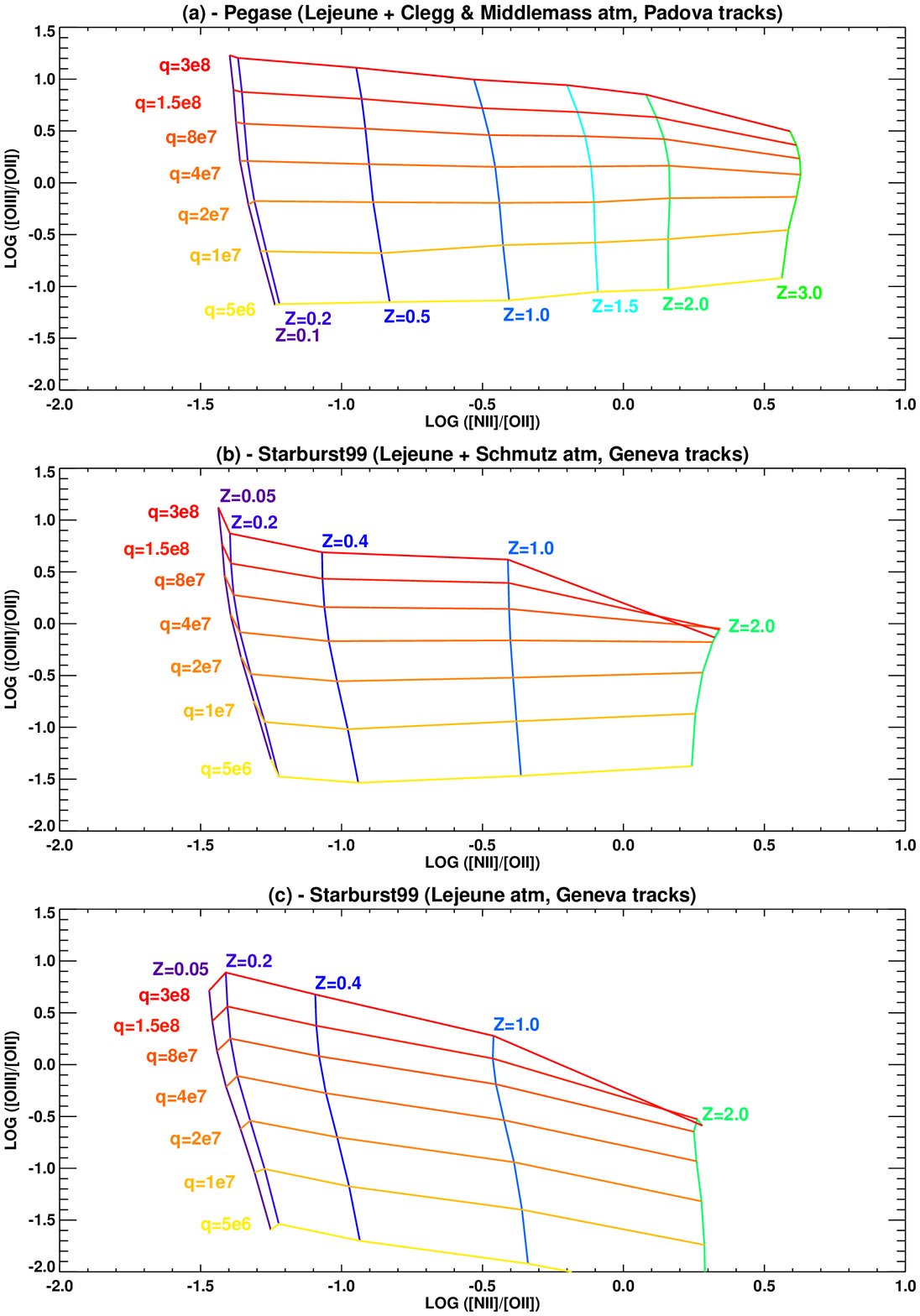}
\figcaption[figure13.ps]{As Figure (9) but for the diagram $\log \mathrm{[NII]/[OII]} $ 
$\emph{vs}$ $\log \mathrm{[OIII]/[OII]}$ We do not have [OII] observations
for the galaxies in our sample, but we provide this diagram for the 
use of the astronomical community. {\label{fig13}} }

\newpage
\epsscale{0.75}
\plotone{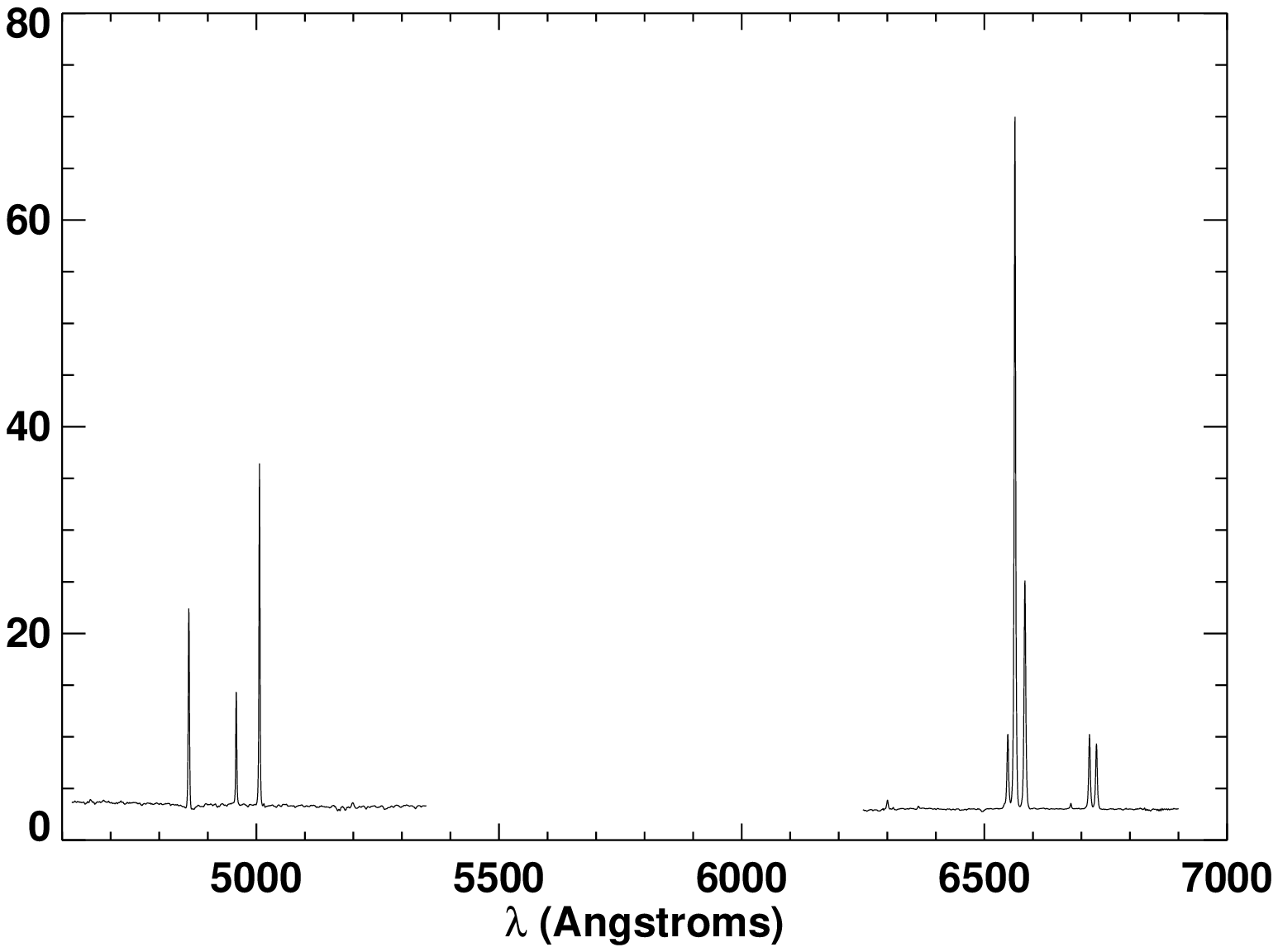}
\figcaption[figure14.ps]{Average spectrum of 56 starbursts in our sample which
have SNRs greater than 60 at H$\beta$, and zero-redshift blue wavelength
cut-off of $\lambda$4620. Flux is in units of $1\times 10^{-15}\,\mathrm{
ergs/s/cm^2/}$\AA {\label{fig14}} }

\plotone{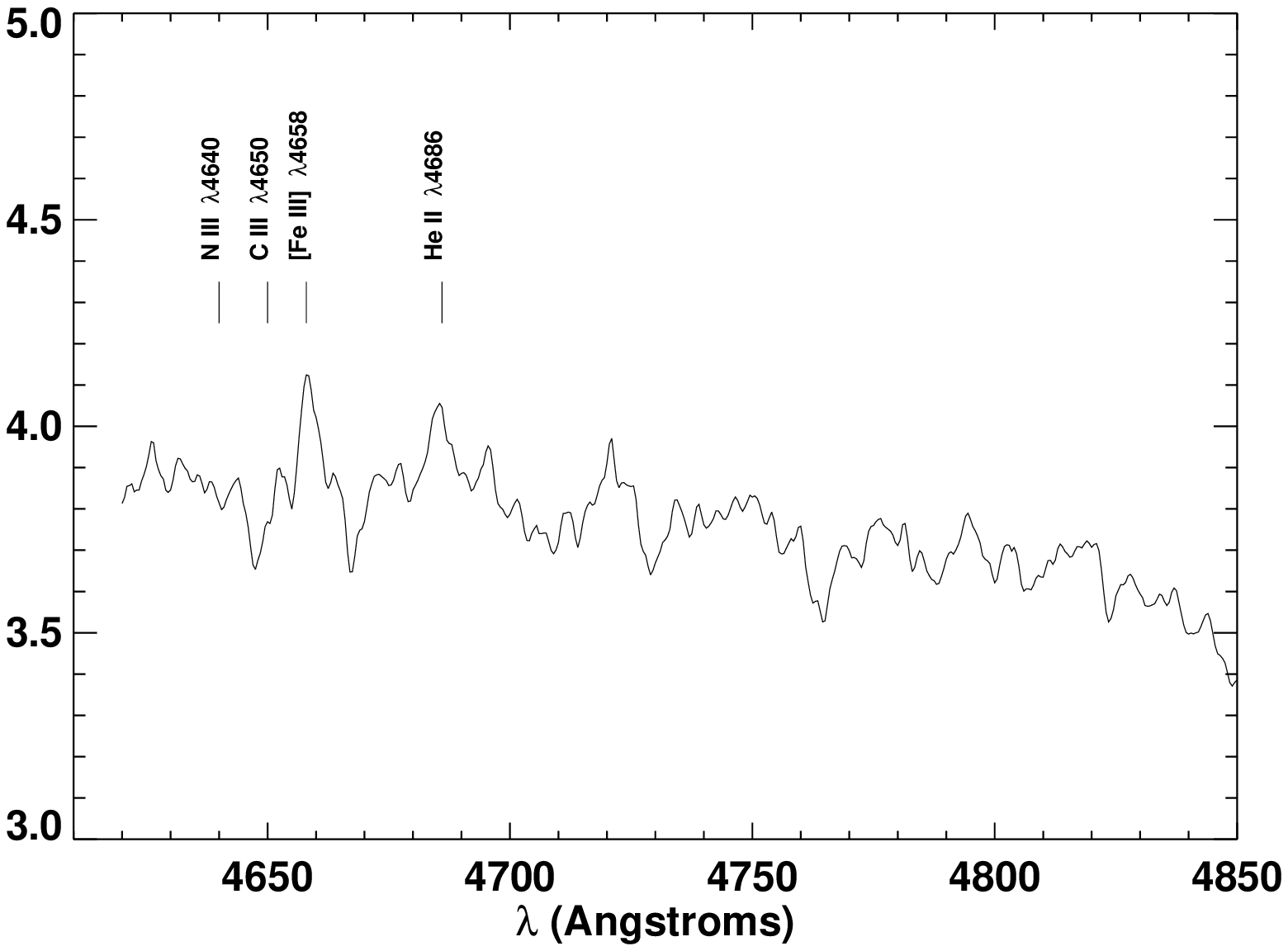}
\figcaption[figure15.ps]{Close-up of Figure (14) showing where one would expect to
see Wolf-Rayet emission features. {\label{fig15}}}

\newpage
\epsscale{0.80}
\plotone{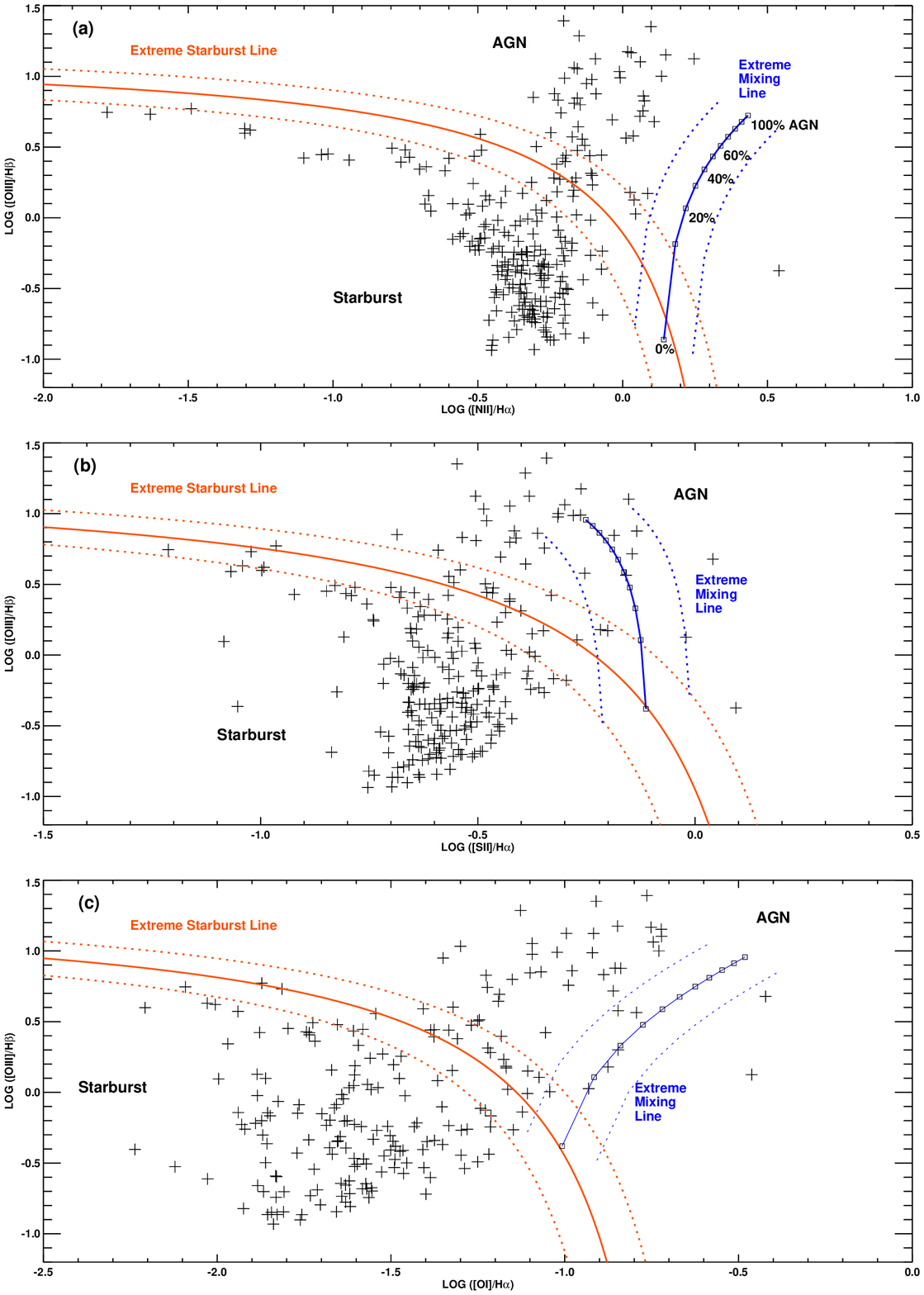}
\figcaption[figure16.ps]{Diagnostic diagrams showing the galaxies in our sample. 
Our theoretical classification line and extreme mixing line are shown in bold, dashed lines
represent $\pm 0.1$ dex of these lines, indicating the error range of our
modeling.  
\label{diagnostics}}

\newpage
\begin{deluxetable}{lrr}
\tabletypesize{\small}
\tablecaption{Solar metallicity ($Z_{\odot}$) and depletion factors (D) adopted
for each element.\label{table1}}
\tablehead{
\colhead{Element}
& \colhead{$\log({\rm Z_{\odot}})$}
& \colhead{$\log({\rm D})$}\\
}
\startdata
H & 0.00 & 0.00 \\
He & -1.01 & 0.00 \\
C & -3.44 & -0.30 \\
N & -3.95 & -0.22 \\
O & -3.07 & -0.22 \\
Ne & -3.91 & 0.00 \\
Mg & -4.42 & -0.70 \\
Si & -4.45 & -1.00 \\
S & -4.79 & 0.00 \\
Ar & -5.44 & 0.00 \\
Ca & -5.64 & -2.52 \\
Fe & -4.33 & -2.00 \\
\enddata
\end{deluxetable}

\begin{deluxetable}{ll}
\tabletypesize{\small}
\tablecaption{Luminosity expected from a 600 km/s shock with a 
spherical precursor of 1pc.\label{table2}}
\tablehead{
\colhead{Species}
& \colhead{L (erg/s)}
}
\startdata
\OIII & $2.5\times10^{39}$ \\
\Hb & $3.3\times10^{38}$  \\
\Ha & $9.8\times10^{40}$ \\

\enddata
\end{deluxetable}

\end{document}